\documentclass[aps,pra,reprint,superscriptaddress,longbibliography]{revtex4-2}

\usepackage{soul}
\usepackage{graphicx}
\usepackage{amsmath,amssymb,amsfonts}
\usepackage{enumitem}
\usepackage{xcolor}
\usepackage{tikz,tkz-euclide}
\usepackage[colorlinks=true,citecolor=blue,linkcolor=blue,urlcolor=blue]{hyperref}
\usepackage[capitalise]{cleveref}
\usepackage{tikz,circuitikz, comment}
\usetikzlibrary{shadows,shadings}
\usepackage{natbib}
\usepackage{braket}
\usepackage[caption=false]{subfig}
\newcommand{\w}{\omega}

\begin{document}

\title{Soliton versus single photon quantum dynamics in arrays of superconducting qubits}
\author{Ben Blain}
\email{ben.blain@tii.ae}
\affiliation{Quantum Research Center, Technology Innovation Institute, Abu Dhabi 9639, UAE}

\author{Giampiero Marchegiani}
\affiliation{Quantum Research Center, Technology Innovation Institute, Abu Dhabi 9639, UAE}

\author{Juan Polo}
\affiliation{Quantum Research Center, Technology Innovation Institute, Abu Dhabi 9639, UAE}

\author{Gianluigi Catelani}
\affiliation{JARA Institute for Quantum Information (PGI-11),Forschungszentrum J\"ulich, 52425 J\"ulich, Germany}
\affiliation{Quantum Research Center, Technology Innovation Institute, Abu Dhabi 9639, UAE}

\author{Luigi~Amico}
\thanks{On leave from the Dipartimento di Fisica e Astronomia ``Ettore Majorana", University of Catania.}
\affiliation{Quantum Research Center, Technology Innovation Institute, Abu Dhabi 9639, UAE}

\affiliation{Centre for Quantum Technologies, National University of Singapore, 3 Science Drive 2, Singapore 117543}
\affiliation{INFN-Sezione di Catania, Via S. Sofia 64, 95127 Catania, Italy}
\affiliation{MajuLab, CNRS-UNS-NUS-NTU International Joint Research Unit, UMI 3654, Singapore}

\begin{abstract}
Superconducting circuits constitute a promising platform for future implementation of quantum processors and simulators. Arrays of capacitively coupled transmon qubits naturally implement the Bose-Hubbard model with attractive on-site interaction. The spectrum of  such many-body  systems is characterized by low-energy localized states defining the lattice analog of bright solitons. Here, 
we demonstrate that these bright solitons can be pinned in the system, and we find that a soliton moves while maintaining its shape. Its velocity obeys a scaling law in terms of the combined interaction and number of constituent bosons. In contrast, the source-to-drain transport of photons through the array occurs through extended states that have higher energy compared to the bright soliton. For weak coupling between the source/drain and the array, the populations of the source and drain oscillate in time;
for chains of even length, their population remains low at all times, while it can reach half the number of total bosons in odd chains. Implications of our results for actual experimental realizations are discussed.
\end{abstract}

\maketitle

\section{Introduction}

Transmons are Josephson junction-based superconducting qubits with reduced sensitivity to charge noise~\cite{PhysRevA.76.042319}.
Networks of transmons are currently being explored for quantum computations, quantum simulations, and quantum sensing applications~\cite{houck2012chip,schmidt2013circuit,lamata2018digital,arute2019quantum,morvan2022formation,kjaergaard2020superconducting,danilin2022quantum}. Specific schemes have recently been proposed, for instance, for the implementation of the two-dimensional Bose-Hubbard model~\cite{Yanay2020Two} and the bosonic quantum East model~\cite{Valencia2022kinetically}. Here we focus on a linear chain of capacitively coupled transmons~\cite{fedorov2021photon}. 
Such systems are generally controlled via microwave transmission lines and resonators. Microwave photons can induce transitions between the transmons' energy levels, and the excitations thus created can propagate through the capacitors as photons. Because of the non-linear inductance of the Josephson junction of the transmon,  these aforementioned  excitations can interact with each other. Overall, the transport of photons through such a non-linear medium can be described in terms of itinerant bosons with a Bose-Hubbard interaction~\cite{hacohen2015cooling}. Recently, Fedorov \textit{et al.}~\cite{fedorov2021photon} have reported photon transport through a transmon chain. In contrast with implementations based on Josephson-junctions arrays (in which the interaction is due to the self-capacitance of the superconducting island)~\cite{fazio2001quantum}, here the  Bose-Hubbard interaction is attractive. Several important problems in quantum science and technology have been studied in Bose-Hubbard as well as more general systems, including driven-dissipative systems~\cite{ma2019A}, many-body localization~\cite{Roushan2017Spectroscopic,Xu2018emulating,PhysRevB.100.134504,Zha2020ergodic,guo2021observation,Gong2021experimental,Chiaro2022direct}, ground state phases in the disorder limit~\cite{Mansikkamaki2021Phases}, correlated quantum walks~\cite{yan2019strongly, Giri2021Two,PainterQRWScience}, and lattice gauge theories~\cite{marcos2013superconducting,Wang2022observation}. 

\begin{figure*} [ht!]
    \includegraphics[width=\linewidth]{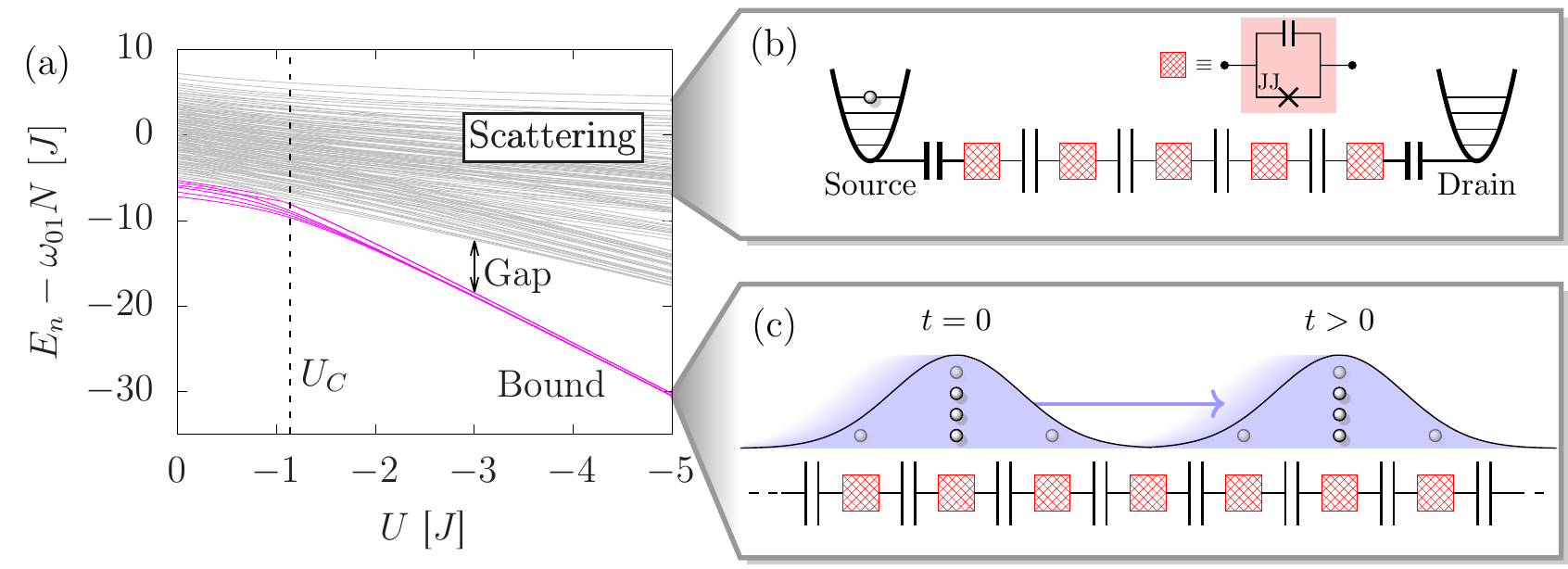}
    \caption{Spectrum and excitation propagation in attractive Bose-Hubbard Hamiltonian. (a) Bose-Hubbard model \cref{eqn:hamiltonianBH} spectrum vs interaction ($U$) for $N=4$ excitations, $M=6$ sites, and no disorder $\mu_i=\w_{01}$. When the interaction is larger than a critical value, $|U|>|U_C|$, a gap in the spectrum forms, separating the bound $N$-particle states from the scattering states. The Bose-Hubbard Hamiltonian is implemented by an array of capacitively-coupled transmon qubits. The effective attractive interaction is determined by the charging energy of the capacitor shunting the Josephson junction (JJ), see the legend for the hatched box in panel (b) representing a single transmon. 
    The different nature of the states in the Bose-Hubbard spectrum is reflected in the dynamics of the excitations. (b) Schematics of source-to-drain excitations transport. A single-mode resonator is prepared in a Fock state and weakly coupled to a transmon chain. Photons  are transported through the chain to a drain resonator, accessing the scattering states associated with single-particle effects. 
    (c) Schematics showing the collective dynamics of localized excitations propagating through a one-dimensional array of capacitively coupled superconducting transmon qubits. The collective motion unveils the nature of the bound state; the ground state of the attractive BH model is a lattice quantum soliton (see also discussion in Sec.~\ref{sec:model}). 
    }
        \label{fig1:schematic}
\end{figure*}

Recently, \citet{SilveriBeyondHardcore} analyzed the transport properties of an array of capacitively coupled transmons beyond the standard hard-core boson treatment. Specifically, they studied the dynamics of localized ``boson stacks'': multi-bosonic excitations localized in the  same site. 
In fact, Bose-Hubbard systems with attractive interactions have been shown to form bound  states defining the analog of bright solitons for  strongly correlated bosonic lattice systems~\cite{naldesi2019rise}. Such solitonic  states are separated from extended states by a characteristic energy gap which increases with the interaction strength, see \cref{fig1:schematic}(a). It turns out that, despite  such bound states being generally less localized compared to boson stacks  (as the particles  spread to nearby lattice sites), they maintain their ``shape''  during their time evolution. 

In this work, we demonstrate how, by a suitable change of a single qubit frequency,  lattice bright solitons can be engineered in a  chain of transmons - see \cref{fig1:schematic}(c). By studying the unitary dynamics, we show that bright solitons can propagate with a remarkable stability, surpassing that of boson stacks. In addition, we analyze the source-to-drain transport of photon excitations injected by resonators - see \cref{fig1:schematic}(b). Such dynamics typically involves single-particle scattering states. Specifically, by matching the frequencies of the source/drain resonators and the qubit (in the array), rather than by solitons, the dynamics results to be dominated by single-excitation transmon states. Nonetheless, by adjusting the frequency detuning between the source-drain and chain, multi-particle transport can be achieved.

This article is organized as follows: the model is introduced in Sec.~\ref{sec:model}. The multi-particle dynamics in a transmon chain is explored in Sec.~\ref{sec:Dynamics}. After discussing the difference between a boson stack and a pinned quantum soliton in Sec.~\ref{sec:stackvspinned}, we study the dependence of the propagation speed of the localized excitations on interaction strength and particle number. The source-to-drain transport is later investigated in Sec.~\ref{sec:sourcetodrain}, where we explore to what degree the localized states of the transmon chain affect the photon transport. In Sec.~\ref{sec:conclusions} we summarise our results.

\section{Model and Methods}
\label{sec:model} 
In a linear network of $M$ capacitively-coupled superconducting transmon qubits, the quantum dynamics of low-energy excitations (plasmons) is governed by the (disordered) Bose-Hubbard~\cite{PhysRev.129.959} Hamiltonian ~\cite{Roushan2017Spectroscopic,PhysRevB.100.134504,fedorov2021photon,hacohen2015cooling,RevModPhys.93.025005}
\begin{equation}
    \mathcal{\hat H}_{BH} = J \sum_{i=1}^{M-1} \left( \hat{b}_{i+1}^\dagger \hat{b}_i + \textrm{h.c.}\right) + \frac{U}{2} \sum_{i=1}^{M} \hat{n}_i (\hat{n}_i - 1) + \sum_{i=1}^{M} \mu_i \hat{n}_i,
\label{eqn:hamiltonianBH}
\end{equation}
where $\hat{b}^\dagger_i$ and $\hat{b}_i$ are bosonic creation and annihilation operators for excitation on site $i$, obeying $\big[ \hat{b}_i, \hat{b}^\dagger_j \big] = \delta_{ij}$. The number operator  $\hat{n}_i = \hat{b}^\dagger_i \hat{b}_i$ counts the number of excitations on the site $i$.

The Hamiltonian of Eq.~\eqref{eqn:hamiltonianBH} consists of three terms. The first term is responsible for the transfer of excitations between nearest-neighbor qubits. The second term gives rise to the attractive on-site interaction, energetically favoring higher occupation of individual transmon qubits. The third term is the on-site chemical potential, which corresponds to the transition frequency between the first two levels of each individual transmon. 

The relations between the Bose-Hubbard parameters entering Eq.~\eqref{eqn:hamiltonianBH} and the physical quantities characterizing the qubits are summarised in~\cref{table:params}. The capacitive couplings between individual transmons in the array give rise to the positive tunneling coefficient $J>0$. Here, for simplicity, we assume that the coupling can be made homogeneous along the array. The interaction between excitations localized on the same transmon corresponds to the transmon anharmonicity, \textit{i.e.}, $U\approx-E_C$, with $E_C$ being the charging energy, also taken homogeneous. Note that, in contrast to  implementations through Josephson-junction arrays~\cite{fazio2001quantum}, the interaction is attractive. The on-site chemical potentials $\mu_i\approx \w_{01}^i$ represent the energy difference between the first two levels of each individual transmon. We note that these chemical potentials can be individually adjusted at each site,  either through fabrication or by using flux-tunable transmons (also called split transmons)~\cite{krantz2019quantum}, making it possible to realize a disordered version of the Bose-Hubbard model.  

The Bose-Hubbard Hamiltonian commutes with the number operator $\hat N=\sum_{i=1}^{M} \hat{n}_i$. In this work, we always consider situations where the total number of excitations in the system
is fixed. As we focus on the Bose-Hubbard description, we will interchangeably use the terms ``particles'' and ``excitations'' when speaking about the transmon excitations in the chain. Therefore, we will also refer to $\hat n_i/N$ as the density operator at site $i$. Due to the weak anharmonicity and the finite height of the cosine potential in a transmon qubit, the transmons are limited in the number of so-called confined excitations (plasmons) they can hold~\cite{lescanne2019escape}. The attractive on-site interaction lowers the energy of multi-particle states. Therefore, the ground state of a system with $N$ excitations will primarily involve Fock states with $N$ excitations in any given site.
\begin{table}
    \begin{tabular}{ccc}
    	\hline
    	\hline
    	Parameter & Description & Physical quantity \\
    	\hline
    	$J$ & Capacitive coupling & $E_C^{ij}/\hbar$ \\
    	$U$ & Anharmonicity & $-E_C^{(i)}/\hbar$ \\
    	$\mu_{i}$ & Transmon frequency & $\omega_{01}^{(i)}$ \\
    	\hline
    \end{tabular}
    \caption{Table of parameters.}
    \label{table:params}
\end{table}

In Fig.~\ref{fig1:schematic}(a), we display the spectrum of the Bose-Hubbard Hamiltonian as a function of the interaction $U$ in the absence of disorder $\mu_i=\w_{01}$, where the last term in Eq.~\eqref{eqn:hamiltonianBH} can be omitted since it gives only a constant shift $\w_{01} N$ to the total spectrum. 
Upon increasing the interaction strength above a critical value $U_C$, the bound states, \textit{i.e.}, the $M$ lowest-energy states, are separated from the scattering states by a finite gap that increases when increasing the magnitude of the interaction $U$. 
This band of bound states is typically referred to as the ``soliton band''~\cite{quantumSoliton1994}.
The ground state of the system corresponds to a superposition of localized bosonic states providing the analog of bright solitons in the quantum regime~\cite{naldesi2019rise} (see also Appendix~\ref{app:state-tomography}).

In this work, we investigate the propagation of a localized bosonic wavepacket through an open-ended chain of transmon qubits. We will consider two distinct protocols (schematically pictured in Figs.~\ref{fig1:schematic}(c) and \ref{fig1:schematic}(b) respectively), exploring both the low-lying and highly excited states of the system. 

\paragraph{ Pin and release dynamics}
\label{par:pin-and-release}
In this protocol, the system is initialized by pinning the bosonic excitations to a specific site, thus selecting a single soliton from the superposition of 
localized states forming the
ground state of the Hamiltonian \cref{eqn:hamiltonianBH}. Such a state can be obtained by setting all of the transmon frequencies to $\mu_i = \omega_{01}$, except for the pinning site $i_\text{pin}$ where  $\mu_{i_\text{pin}}=\w_{01}-\mu_\text{pin}$. The value of $\mu_\text{pin}>0$ is chosen~\cite{naldesi2019rise} so that a single localized quantum soliton is projected over the overall localized ground state of the Hamiltonian with homogeneous chemical potential.
This is achieved by setting the pinning strength to exactly the width of the solitonic energy band, \textit{i.e.}, $\mu_\textrm{pin}=\mu_{\rm band}(U,N)$, as expressed by the condition~\cite{naldesi2019rise}
\begin{equation}
    \mu_\text{band}(U,N) =  \frac{|U| (N-1)}{2} \left(\sqrt{1+\frac{16J^2}{U^2 (N-1)^2}}-1\right).
    \label{eqn:mupin}
\end{equation}
Depending on the tunable disorder landscape, the pinned soliton may symmetrically propagate towards both edges of the chain [discussed below in Sec.~\ref{sec:stackvspinned}] or move preferentially in a selected direction [investigated in Sec.~\ref{sec:directional}], with the latter being the possibility displayed in Fig.~\ref{fig1:schematic}(c).

\paragraph{Source-to-drain photon transport}

In this scheme, we consider a typical experimental setup~\cite{fedorov2021photon}, where a resonator is attached to each end of the transmon chain, below denoted as the source and the drain. In this configuration, we study the source-to-drain transport of electromagnetic excitations, prepared in the source as non-interacting photons, and mediated by quantized plasma oscillations while traveling through the transmon array. The Hamiltonian of this system is $\hat{\mathcal{H}}=\hat{\mathcal{H}}_{\rm BH}+\hat{\mathcal{H}}_{\rm SD}$, where 
\begin{equation}
    \mathcal{\hat H}_\text{SD} = \sum_{\alpha=S,D} \w_\alpha \hat{a}^\dagger_\alpha \hat{a}_\alpha + J' \left( \hat{a}_S\hat{b}_{1}^\dagger  + \hat{a}_{D}\hat{b}_{M}^\dagger  + \textrm{h.c.} \right) 
    \label{eqn:H_SD}
\end{equation}
Here, $\w_{S(D)}$ is the resonating frequency of the source (drain) resonator~\footnote{We consider a single-mode resonator in this work.}, and $\hat{a}_{S(D)},\hat{a}^\dagger_{S(D)}$ are the corresponding annihilation/creation operators. 
In this protocol, we first prepare the source resonator in a Fock state, and then we connect the source and drain oscillators to the main system by a quench of the coupling $J'$ from zero to a finite value. The excitations, initially prepared as non-interacting photons in the source resonator, enter the transmon chain through the capacitive couplings. Then the excitations in the first transmon are annihilated as the next transmon is excited, and so on, thus carrying the excitation energy through the system. For the purposes of studying the source-to-drain transport dynamics, we consider the source and drain to be resonant $\w_S = \w_D=\w_r$, and equal-frequency transmons in the chain $\mu_i=\w_{01}$. Depending on the detuning between the source (drain) and the transmon frequencies $\Delta=\w_r-\w_{01}$, the photon energy in the source (drain) spectrum may resonate with different energy levels within the central chain. We note that in this case, the Hamiltonian $\hat{\mathcal{H}}=\hat{\mathcal{H}}_{\rm BH}+\hat{\mathcal{H}}_{\rm SD}$ commutes with the total number of excitations $\hat{\mathcal{N}} =\hat{N}+\sum_\alpha\hat{a}^\dagger_\alpha \hat{a}_\alpha$, which in this protocol is fixed by the number of photons initially in the source.

All the numerics displayed in this work are obtained through exact diagonalization~\cite{lin1990exact,zhang2010exact}. In particular, we always consider a closed system evolution characterized by a fixed number of excitations (either $N$ or $\mathcal N$ in protocols \textit{a} and \textit{b}, respectively), neglecting particle losses (that is, relaxation) and other forms of decoherence. Throughout this work, $\langle \dots \rangle$ denotes the expectation value over the time evolved state $|\psi(t)\rangle = e^{-\imath \hat{\mathcal{H}} t} |\psi_0\rangle$. In specific limits, we provide analytical approximation derived with specific techniques mainly discussed in the Appendices. In the simulations, and below, we set the reduced Planck constant to unity, \textit{i.e.}, $\hbar = 1$. 
\begin{figure*}
    \centering
    \includegraphics[width=\linewidth]{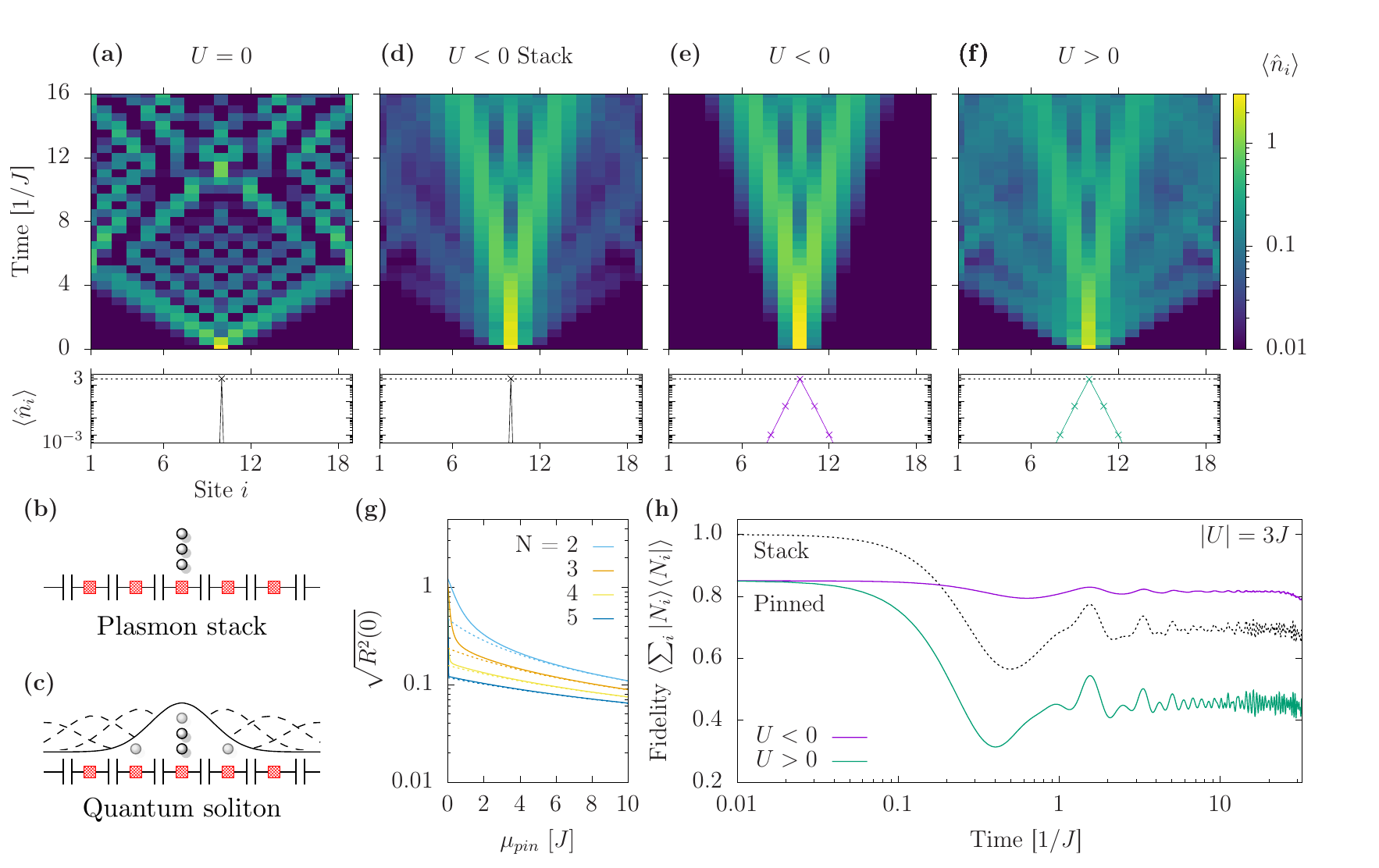}
    
    \caption{Propagation of localized bosonic excitations. (a,d-f) The time evolution of the site occupation number $\langle \hat{n}_i\rangle$ of different initial states of a system of $M=19$ sites and $N=3$ particles, for (a) zero (equivalent to $N=1$), (d-e) attractive ($U = -3J$) and (f) repulsive ($U = +3J$) interaction. In (a) and (d) a stack of 3 plasmons is initially placed on the central qubit. Else, a quantum soliton has been placed in center of the system through pinning $\mu_{\rm pin}=\mu_{\rm band}(U,N)$ [Eq.~\eqref{eqn:mupin}] in {(e)}. Note the zero-time cross-section below the density plot, showing the initial density: the occupation of the central site is slightly smaller than $3$ in (e), and the remaining fraction occupies the neighboring sites. In the repulsive case (f), we consider the the same initial state as in panel (e). Panels (b) and (c) schematically mimic the initial density distribution of a boson (plasmon) stack and a pinned quantum soliton (solid curve), respectively. The boson stack has zero width, while a quantum soliton occupies neighboring sites. The dashed collection of bell curves in (c),  modulated by an envelope, schematically represents the ground state of the unpinned Bose-Hubbard Hamiltonian. (g) The width of a pinned soliton $\sqrt{R^2(0)}$ (see definition in Sec.~\ref{sec:stackvspinned}) vs pinning strength $\mu_\text{pin}$, for various particle number $N$ and $U=-3J$. In the limit of infinite pinning ($\mu_\text{pin} \to +\infty$), the width of the pinned density distribution tends to zero, reproducing a boson-stack. Dashed curves are the analytical approximations for large pinning (derived in Appendix~\ref{app:soliton-width-derivation}), as discussed in Sec.~\ref{sec:stackvspinned}. 
    (h) Dynamics of the expectation value of the projector over the subspace generated by the boson stack states. The evolution of a  boson-stack (dashed) is compared with the initially-pinned quantum soliton (solid). For the quantum soliton, the evolution depends on the sign of the interaction, in agreement with the plots in panels (e) and (f).
    }
    \label{fig2:bosonvsstack}
\end{figure*}

\section{Dynamics of localized multi-boson excitations}
\label{sec:Dynamics}
The interacting nature of the plasmonic excitations in the transmon chain strongly influences
the system dynamics when the excitations are localized on a single site. This situation corresponds to having a particular transmon prepared in a highly excited state~\cite{peterer2015coherence,SilveriBeyondHardcore}.
As a reference to the general cases, we first consider the time evolution of a single excitation initially placed in the center of the lattice by means of exciting the central transmon qubit to the first excited state. For a single excitation, the time evolution is independent of the interaction $U$ and equivalent to the density dynamics of a non-interacting system ($U=0$). To visualize the single-excitation evolution, and still comparing with the interacting cases discussed below, we show in Fig.~\ref{fig2:bosonvsstack}(a) the dynamics of three non-interacting excitations.

Due to the capacitive coupling, the excitations evolve symmetrically towards both edges of the chain, where they are reflected. The chessboard-like pattern in the density evolution results from self-interference due to quantum superposition.
The evolution is similar to a continuous-time quantum random walk on a lattice~\cite{Giri2021Two,Melnikov2016Quantum,Lahini2018Quantum,Beggi2018Probing,Das_2022}, in that, with time evolution, the excitations have an equal probability of moving to the left or the right of their current position. We remark that the interference pattern for a single excitation is also well-understood on the experimental side in one and two-dimensional transmon arrays~\cite{Ye2019propagation,Gong2021experimental,karamlou2022quantum,Zhao2022probing,braumuller2022probing}.
\subsection{Boson stack vs. pinned soliton}
\label{sec:stackvspinned}
The ground state of a Bose-Hubbard Hamiltonian with attractive interaction and $N$ excitations is a delocalized superposition of quantum solitons centered on each site of the chain, with an overall envelope determined by the finite size of the chain 
[see Fig.~\ref{fig2:bosonvsstack}(c) and Appendix~\ref{app:state-tomography}]. 
\begin{figure*}[htp]
    \centering
    \includegraphics[width=\linewidth]{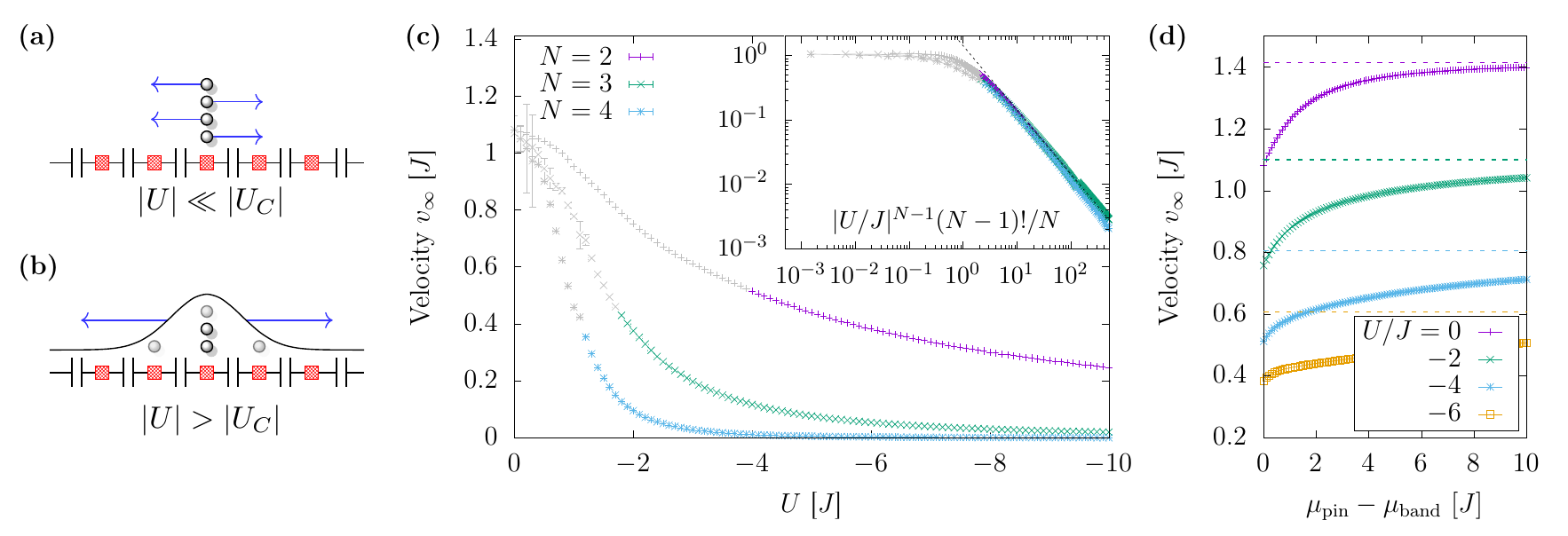}
    
    \caption{Expansion velocity of a pinned quantum soliton. (a-b) Schematics showing two extreme dynamical regimes: (a) weakly-interacting $|U|\ll |U_C|$, where excitations sequentially tunnel, and (b) strongly interacting ($|U|\geq |U_C|$), where all of the excitations  move collectively as a composite particle. 
    (c) Asymptotic expansion velocity $v_\infty$ (for an infinite chain, see text) vs on-site interaction $U$ of a pinned soliton $\mu_{\rm pin}=\mu_{\rm band}(U,N)$ [Eq.~\eqref{eqn:mupin}]
    for different particle numbers $N$ and $M=29$ sites. The interaction effectively increases the inertia of the localized excitations.  
    Grey points represent values where there is no gap in the spectrum between the scattering and bound states~\cite{naldesi2019rise}.  
    In the inset, we show the collapse of the velocity curves for different $N$, using the scaling $|U/J|^{N-1}(N-1)!/N$, 
    derived by \citet{SilveriBeyondHardcore} for $|U|\gg J$. The dashed line in the inset shows the velocity of a boson stack that is formed at very large values of the on-site interaction; the stack behaves effectively as single particle~\cite{SilveriBeyondHardcore} and its velocity is that of a quantum random walk with renormalized tunneling $J\to \tilde J$ (see main text). 
    (d) The expansion velocity as a function of pinning shift $\mu_\text{pin}-\mu_\text{band}$ for $N=2$ and various values of on-site interaction, where $\mu_\text{band}$ [\cref{eqn:mupin}] is the value of $\mu_\text{pin}$ to form a bright soliton. The dashed lines show the velocity of a boson stack, which is equivalent to the pinned soliton in the limit $\mu_\text{pin}\to\infty$.
  }
\label{fig3:velocity}
\end{figure*}

When addressing the multi-excitation dynamics in a transmon chain, different scenarios can be investigated. First, in full analogy with the single-excitation case, the central-site transmon can be prepared in the $N$-boson state with a series of $\pi$-pulses~\cite{peterer2015coherence}, forming a stack [see Fig.~\ref{fig2:bosonvsstack}(b)], and then it is let to evolve. This situation has recently been theoretically investigated in Ref.~\cite{SilveriBeyondHardcore} in the strongly interacting limit, and the site-density evolution is displayed in Fig.~\ref{fig2:bosonvsstack}(d). Conversely, the transmon frequency in the central site can be detuned, setting $\mu_{i_{\rm pin}}=\w_{01}-\mu_{\rm pin}$, to localize the excitations. In this case, the ground state of the pinned $N$-particle Bose-Hubbard Hamiltonian is characterized by a pinned soliton, as schematically represented by the solid curve in Fig.~\ref{fig2:bosonvsstack}(c). The dynamics is triggered by removing the frequency detuning (pinning), leading to the density evolution in Fig.~\ref{fig2:bosonvsstack}(e). The tunable detuning can be realized, for instance, with a flux line controlling a split transmon~\cite{PhysRevA.76.042319}. In both protocols, the excitations are localized in the central site for a significantly longer time with respect to the non-interacting case. Yet, the boson stack shows additional low-density components which resemble the single-particle propagation. We remark that these differences are the results of small differences in the initial state $\ket{\psi_0}$, namely: i) $\ket{\psi_0}=(\hat{b}^{\dagger}_{i_{\rm pin}})^N\ket{0}/\sqrt{N!}$, with $\ket{0}$ denoting the vacuum state, for the boson stack; ii) $\ket{\psi_0}$ is the ground state of the pinned Hamiltonian (the lowest eigenstate obtained by exact diagonalization of the Bose-Hubbard Hamiltonian \cref{eqn:hamiltonianBH} with the pinning potential $\mu_{i_{\rm pin}}=\w_{01}-\mu_{\rm pin}$) for the soliton case -- see also Appendix~\ref{app:state-tomography}. In the plot at the bottom of Fig.~\ref{fig2:bosonvsstack}(e), we show how the pinned soliton has a non-zero density occupation at sites adjacent to the pinning center, with a characteristic exponential decay~\cite{naldesi2019rise}. We characterize the spread $R(t)$ of the soliton at time $t$ around the pining site by introducing the quantity $R^2(t) = \frac{1}{N}\sum_{i=1}^M \langle \hat n_i(t)\rangle (i-i_\text{pin})^2$. 

In~\cref{fig2:bosonvsstack}(g), we display the initial size of the pinned soliton $\sqrt{R^2(0)}$ as a function of the pinning strength for different $N$. For large pinning strength $\mu_\text{pin}\gg J$, the width of a pinned quantum soliton is approximately expressed as 
$
\sqrt{R^2(0)} \approx \sqrt{2} J/[\mu_\text{pin} + |U| (N-1)]
$, as shown in dashed lines in Fig.~\ref{fig2:bosonvsstack}(g) (see \cref{app:soliton-width-derivation}).

To stress the difference between the boson stack and the pinned soliton, we estimate the $N$-particle component of the time-evolved state~\footnote{We recall that ``particle'' is used here to indicate the bosonic excitation.}. This feature is highlighted in \cref{fig2:bosonvsstack}(h), where we display the expectation value of the projector on the $N$-particle state. In other words, we compute the expectation value of the projector $P^{(N)}=\sum_{i=1}^M\ket{N_i}\bra{N_i}$ onto the subspace generated by the states with $N$ excitations in site $i$ (here denoted with  the short-hand $\ket{N_i}$), 
\begin{equation}
\mathcal{F}_N(t)\equiv \braket{P^{(N)}}=\sum_{i=1}^M |\braket{\psi(t)|N_i}|^2.
\label{eqn:fidelity}
\end{equation}
Note that, by definition, this fidelity is smaller than or equal to one. We introduce this quantity as a figure of merit of the $N$-particle component of the state at arbitrary time $t$; indeed the bound states [cf. Fig.~\ref{fig1:schematic}(a)] in the Bose-Hubbard Hamiltonian for $|U|>|U_C|$ are mainly composed of a linear superposition of the boson-stack states $\ket{N_i}$~\footnote{We note that for small chains, the ground state also has a sizeable overlap with states $\ket{(N-1)_{i_{\rm pin}}, 1_{i_{\rm pin}\pm 1}}$, 
using the notation $\ket{N_{i},N_{j}} = \left(\hat{b}^\dagger_i\right)^{N_i}\left(\hat{b}^\dagger_j\right)^{N_j} \ket{0}/\sqrt{N_i!N_j!}$; however, these overlaps are strongly suppressed for larger system sizes.}. We classify as \textit{stable} a state for which the fidelity does not significantly change with time. For the boson stack evolution of \cref{fig2:bosonvsstack}(h) (dashed black curve), this fidelity is one at $t=0$, but decays on the typical single-particle tunneling time $\approx 1/J$, and oscillates around 0.7. On the other hand, the fidelity of a pinned soliton (solid) is smaller than the stack at $t=0$, but for negative $U$ it remains almost unaffected during the time evolution. 
That is, the dynamics is mostly supported by $N$-particle states and the single particle ones do not significantly contribute to
the evolution. 
Upon increasing the strength of the attractive interaction, \textit{e.g.}, $U=-10 J$, the differences between a pinned soliton and a boson stack get smaller, and the fidelity is closer to 1 (not shown). In this respect, the boson stack is the limiting state for a pinned soliton in the infinite interaction limit $-U\gg J$, or, similarly, in the infinite pinning limit $\mu_\mathrm{pin}\to\infty$ for a given attractive interaction. Yet, in general, we observe that the pinned soliton shows stronger stability. It is important to note that, as $\mu_i$ represents the transition frequency of a transmon, this must be positive and therefore $\mu_\mathrm{pin} < \omega_{01}$. 

To conclude this section, we briefly comment on the dynamics for repulsive interactions, $U>0$. When multiple excitations are considered in the chain $N\geq 2$, the sign of the interaction term in the Bose-Hubbard Hamiltonian can be probed, for instance, through the two-site correlations~\cite{Beggi2018Probing}. Here, we emphasize how the site-density evolution can highlight the nature of the ground state. Upon initializing the state as in the pinned soliton case for $U<0$, we quench the interactions to a positive value~\footnote{The quench may be performed for a two-excitation state in a C-shunt flux qubit, switching the magnetic flux from zero to half a flux quantum.} and obtain the evolution in Fig.~\ref{fig2:bosonvsstack}(f). Due to the repulsive interaction, the single-particle-driven dynamics is strongly enhanced with respect to the attractive case, although few multiparticle effects are still present due to the high-energy bound states that exist in the repulsive case~\cite{wang2010preparation}. This view is confirmed in the fidelity plot [green curve in Fig.~\ref{fig2:bosonvsstack}(h)], where the time-evolved state retains a multiparticle contribution of roughly 0.4, significantly lower than for $U<0$ due to the different nature of the ground state. 
Finally, we numerically verify that for a boson stack, the dynamics is independent of the sign of $U$. In this respect, the boson stack and the pinned soliton show quite different behavior, which can be understood in terms of the $U\leftrightarrow -U$ inversion theorem~\cite{winkler2006repulsively,ronzheimer2013expansion}. Indeed, this result, stating that some operators, such as the density, are independent of the sign of the interaction, applies only to specific initial states~\cite{Boschi2014bound}, including the boson stack but not the highly-entangled pinned soliton state.

\subsection{Expansion velocity}
\label{sec:velocity}
In the previous subsection, we discussed the site-density evolution for different values of the interactions, and highlighted the difference between a $N$-bosons stack and the state realized through pinning, which leads to a quantum soliton. Here we want to quantitatively address the speed of the excitation propagation as a function of the interaction. For a noninteracting chain $U=0$, the evolution of the single-particle excitation placed on a specific site can be mapped to a continuous quantum walk~\cite{Das_2022}, where the site-density occupation represents the probability. It is well known that the quantum walk propagation is ballistic~\cite{karamlou2022quantum}, and so characterized by a linear-in-time root mean square deviation (RMSD) $\sqrt{R^2(t)}\propto t$. As a consequence, it is possible to define the spread velocity $v\approx\sqrt{R^2(t)}/t$. In this present case, the velocity definition has a few complications. Firstly, for a finite size system, the relation $\sqrt{R^2(t)}\propto t$ only holds on a short timescale $t\lesssim M/2v$, even at $U=0$. Moreover, the interaction term significantly modifies the time evolution, possibly spoiling the linear-time scaling.
Following Refs.~\cite{ronzheimer2013expansion,Boschi2014bound,naldesi2019rise}, we define the expansion velocity as 
\begin{equation}
    \label{eqn:velocity}
    v(t) = \frac{d}{dt} \sqrt{R^2(t) - R^2(0)}.
\end{equation}
This definition is known to be quite robust against finite-size effects on a short timescale (see~\cite{ronzheimer2013expansion} and \cref{appendix:steady-state}).
In general, this velocity is a function of time, so we consider the asymptotic value $v_\infty$, following the fitting procedure detailed in \cref{appendix:steady-state} and discussed in Ref.~\cite{Boschi2014bound,naldesi2019rise}~\footnote{We also verified that the results are robust to the specific method used in the analysis; for instance, the curves are not significantly affected if the fitting is replaced by a time-averaging procedure.}.

The fitted velocity $v_\infty$ is plotted as a function of $U$ for different values of $N$ in Fig.~\ref{fig3:velocity}(c). The velocity is a monotonically decreasing function of both $U$ and $N$, due to the increasing interaction energy. For $|U|\ll J$, the velocity is independent of the excitation number, where the density-wave transport is dominated by single-particle effects [see the schematic Fig.~\ref{fig3:velocity}(a)]. We find that the velocity depends on a suitable combination of $U$ and $N$, displaying certain traits of universality, see the inset of \cref{fig3:velocity}(c). In particular, the velocity reduction is related to the effective rescaling of the tunneling term $J\to \tilde J$, where $\tilde J=JN(J/|U|)^{N-1}/(N-1)!$ represents the tunneling probability for an $N$-boson particle in the limit $|U|\gg J$, as recently discussed in Ref.~\cite{SilveriBeyondHardcore}. In other words, for strong interactions $|U|\gg J$, the excitations' propagation occurs as if the quantum soliton were a collective particle, see Fig.~\ref{fig3:velocity}(b). Indeed, in this limit, the pinned soliton is equivalent to a boson stack, as discussed in Sec.~\ref{sec:stackvspinned}; the propagation speed approximately reads  $v\approx\sqrt{2}\tilde{J}$, see the black dashed line in the inset of Fig.~\ref{fig3:velocity}(c). Notably, the scaling  $v\propto\tilde J$ was identified as the group velocity of a quantum lattice soliton in the strong-interaction limit even earlier~\cite{quantumSoliton1994}. We also note that these results are in good agreement with the $v_\infty\propto 1/|U|$ scaling numerically obtained in Ref.~\cite{Boschi2014bound} for $N=2$ and large $|U|$~\footnote{In Ref.~\cite{Boschi2014bound}, the authors investigate repulsive interactions $U>0$; yet, for a boson stack, the inversion theorem $U\to -U$ holds, making the expectation value of the number operators independent on the sign of $U$, see Appendix B of Ref.~\cite{Boschi2014bound}.}, and that the collapse of the colored symbols in the inset of \cref{fig3:velocity}(c) suggests a way to estimate the critical interaction strength $U_C$, see Appendix~\ref{app:UC}.

In closing this section, we briefly comment on the impact of the pinning strength $\mu_{\rm pin}$ on the expansion velocity. In \cref{fig3:velocity}(d), we plot $v_{\infty}$ as a function of $\mu_{\rm pin}$ (larger than the optimal value $\mu_{\rm band}$) for $N=2$ and different values of the interaction. The expansion velocity increases monotonically with the pinning strength for every value of $U$. In the limit of infinite pinning $\mu_{\rm pin}\to\infty$, $v_\infty$ saturates to the expansion velocity of a boson stack, as shown with dashed lines in \cref{fig3:velocity}(d) for each value of $U$. We note that in the non-interacting limit $U\to 0$, this velocity reads $v_\infty\approx\sqrt{2}J$, as well-known in the case of the standard quantum walk~\cite{Boschi2014bound,karamlou2022quantum}, while for large interaction ($|U|\geq |U_C|$) the velocity saturates to a value typically larger than $\sqrt{2}\tilde{J}$ (not shown here), probably due to the effect of the scattering states, which transport excitations at a faster rate [cf. the outer-edge components in Fig.\ref{fig2:bosonvsstack}(d)].

\subsection{Directional transport}
\label{sec:directional}
\begin{figure*}
    \centering
    \includegraphics[width=\linewidth]{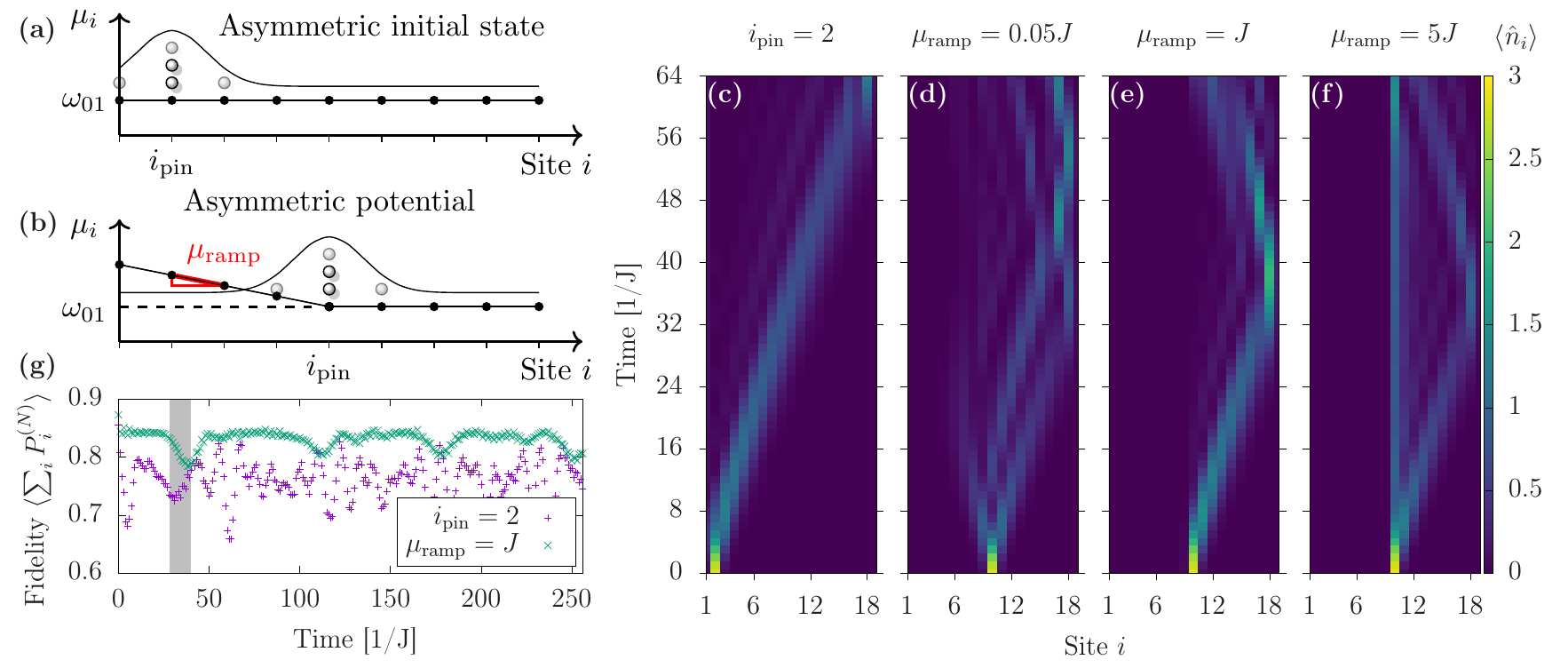}
    
    \caption{Directional transport of a quantum soliton. 
    (a)-(b) Schematic diagrams showing two different protocols for directional propagation. 
    (a) Asymmetric initial state. The soliton is pinned at site $2$, and left to evolve afterwards, moving primarily towards a higher site-number. 
    (b) Asymmetric potential. The soliton is pinned at the central site $i_{\rm pin}=(M+1)/2$, while the chemical potential landscape is increased monotonically for $i<i_{\rm pin}$, giving directionality to the time-evolved state after releasing the pinning. 
    (c)-(f) Dynamics of the site occupation number $\langle\hat{n}_i(t)\rangle$ after pinning and releasing, for (c) site-2 pinning and no ramping, and (d)-(f) central-site pinning and different ramp heights.  
   (g) Fidelity with the subspace generated by the $N$-particle localized states for the two protocols, corresponding to the dynamics displayed in Figs.~\ref{fig4:ramp-heights}(c) and \ref{fig4:ramp-heights}(e). 
    The shaded region approximately shows the time where the boundary ($i=M-1$) is first reached by the traveling quantum soliton for both protocols. Parameters are $N=3$, $M=19$, and  $U = -3J$, except in (g) where, for $i_\text{pin}=2$, we use $M=9$ to have the same distance between the effective boundaries in the two considered cases. In all the panels, the soliton is pinned with the value discussed in Sec.~\ref{sec:stackvspinned}, \textit{i.e.}, $\mu_{\rm pin}=\mu_{\rm band}(U,N)$ [Eq.~\eqref{eqn:mupin}].
    }
    \label{fig4:ramp-heights}
\end{figure*}

Above, we discussed the spreading of the quantum soliton when pinned at the center of the transmon array. Due to symmetry, the excitations  propagate toward both edges of the chain equally. Here, we discuss simple protocols in which the excitations move preferentially in one direction. 
To achieve this goal, the left-right symmetry with respect to the central site in the chain must be broken. This can be done by either i) preparing a left-right asymmetric initial state (near-edge propagation) as shown in Fig.~\ref{fig4:ramp-heights}(a), or ii) considering an asymmetric potential landscape (as given by the on-site chemical potential $\mu_i$), see Fig.~\ref{fig4:ramp-heights}(b).

The first case is obtained by changing the pinning site, say, setting $i_\text{pin}=2$~. We note that by selecting $i_\text{pin}=1$ the propagation of the excitations is strongly suppressed due to edge localization, a phenomenon discussed in Refs.~\cite{Pinto,SilveriBeyondHardcore,footnote1}. 
The site-density dynamics is shown in~\cref{fig4:ramp-heights}(c). The peak in the density profile shifts towards increasing site-number indices due to the initial asymmetry. Differently from the situation where the soliton is pinned in the central site, the spreading of a soliton does not occur in the same measure, but, rather, excitations moving to the left of the pinning site are reflected by the boundary, thus following the density peak, moving to the right. As a consequence, the excitations remain spatially localized about the site-index of the density peak, effectively approximately retaining the initial width of a pinned soliton.
While reaching the rightmost boundary [site 19 in Fig.~\ref{fig4:ramp-heights}(c)], the excitations which initially moved left and right constructively interfere [for $t\approx 60/J$ in Fig.~\ref{fig4:ramp-heights}(c)] and the density at the peak is increased again. In starting as close to a boundary as possible, with $i_\text{pin}=2$, the excitations initially moving left are reflected in the shortest time, keeping the effective width of the density distribution as narrow as possible about the peak density site. We note that this protocol is less effective at transporting a spatially-localized bright soliton when $i_\text{pin}>2$ (not shown).

The second protocol is implemented, for instance, by considering a ``ramp''-like potential landscape, see Fig.~\ref{fig4:ramp-heights}(b), achieved by individually tuning the transmon frequencies
\begin{align}
\mu_i &= \omega_{01}+\mu_{\rm ramp}(i_{\rm ramp}-i) &i< i_{\rm ramp}\nonumber\\
\mu_i &= \omega_{01} &i\geq i_{\rm ramp}
\label{eqn:h-ramp}
\end{align}
where $i_{\rm ramp}$ is the site at the base of the ``ramp'', and $\mu_{\rm ramp}>0$ is the frequency detuning between adjacent transmons in the ramp.

We now consider the propagation of a pinned soliton for different values of $\mu_{\rm ramp}$, fixing $i_{\rm ramp}=(M+1)/2$. As in the protocol discussed previously, the pinning potential is removed at $t=0$. The ramp effectively implements a soft boundary; the excitations are progressively reflected while increasing the potential step $\mu_{\rm ramp}$, and consequently depleting the excitations in the region $i<i_{\rm ramp}$. 
For a small ramped potential $\mu_{\rm ramp}\ll J$, 
the plasmonic excitations are gently pushed in the direction where the frequency detuning $\mu_i-\w_{01}$ is zero, as seen in Fig.~\ref{fig4:ramp-heights}(d). In the evolution, the site-density displays multiple peaks due to the low value of the ramping, while still showing asymmetric evolution. As in the protocol where the excitations are localized to the second site, the excitations partially recombine upon reaching the opposite end, although never reaching the initial peak density of the pinned soliton. By increasing the ramp strength $\mu_{\rm ramp}=J$ [see~\cref{fig4:ramp-heights}(e)], the excitations only propagate in the region $i\geq i_{\rm ramp}$, being delocalized on a size approximately equal to the one of the initially pinned soliton. The excitations become more delocalized after reflection from the right-boundary, due to interference effects. For a large value of the ramp $\mu_{\rm ramp}=5 J$, as shown in Fig.~\ref{fig4:ramp-heights}(f), the ramp effectively acts as a boundary. As a result, a finite excitation density remains trapped in the pinning site, which is reminiscent of edge localization~\cite{SilveriBeyondHardcore, Pinto}. These results indicate that there exists an optimal ramp slope
to impose directionality on the evolution of the pinned excitations, since the soliton is symmetrically spreading for $\mu_{\rm ramp}/J\to 0$ and stationary for $\mu_{\rm ramp}/J\to +\infty$ (limit not shown here). 

In closing this section, we compare the $N$-particle component (see definition in Sec.~\ref{sec:stackvspinned}) as a function of time for the two protocols in Figs.~\ref{fig4:ramp-heights}(a) and \ref{fig4:ramp-heights}(b), in analogy with the analysis performed for the soliton pinned at the center of the chain [see Sec.~\ref{sec:stackvspinned} and Fig.~\ref{fig2:bosonvsstack}(h)]. Notably, the presence of the ramp significantly improves the stability of the pinned soliton with respect to the near-edge preparation; the fidelity with the $N$-particle projected state remains approximately constant in the evolution, even after multiple reflections from the hard boundary, except for the particular times where the excitations are being reflected at the rightmost edge or at the base of the ramp. 

\begin{figure*}
    
    \centering
    \includegraphics[width=\textwidth]{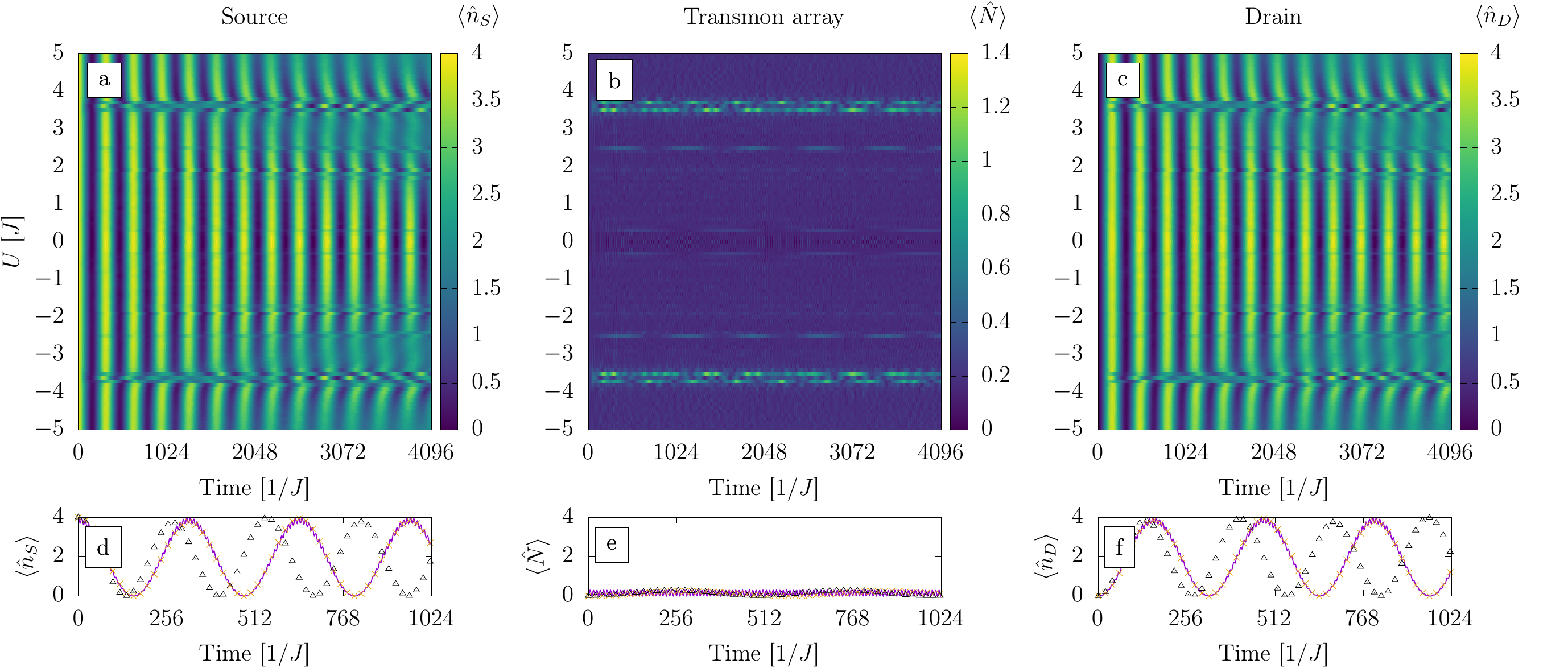}
    \includegraphics[width=\textwidth]{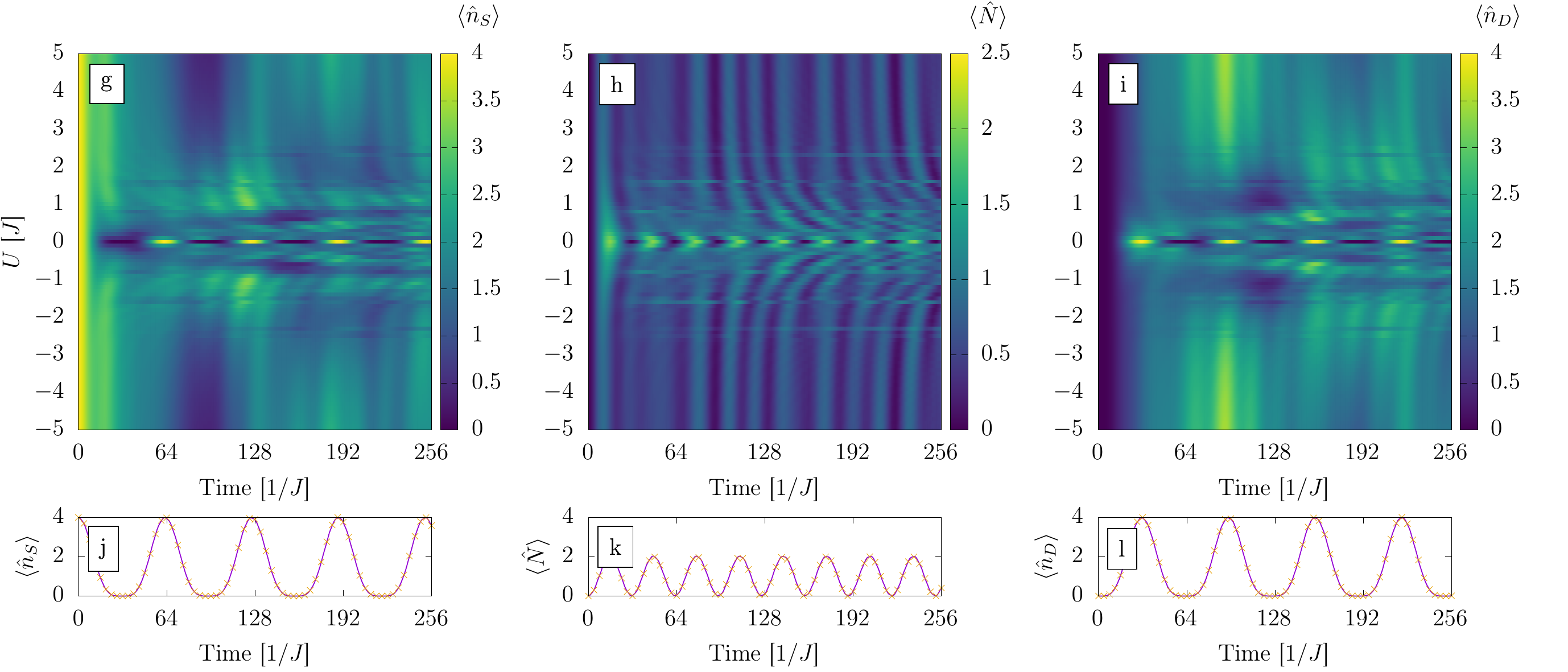}
    
    \caption{Source-to-drain photon transport in the resonant case between source/drain and transmon qubits ($\w_r=\w_{01}$). Time evolution of the occupation numbers of the (left) source, (center) chain, and (right) drain for an (a)-(f) even ($M=4$) and (g)-(l) odd ($M=3$) number of sites. The source and drain resonators are weakly coupled to the transmon array, with $J' = 0.1J$. (d)-(f),(j)-(l) Comparison between the numerics (lines) and the analytics approximations (points) for the $U=0$ cross-section, (d)-(f): \cref{eqn:density-source-U0-even,eqn:density-drain-U0-even}, (j)-(l): \cref{eqn:density-source-U0,eqn:density-drain-U0} derived in Appendix~\ref{app:odd-chain}. The triangles in (d)-(f) show the analytics 
    when not accounting for level repulsion in the even parity case, see Appendix~\ref{app:even-chain}. The dynamics is independent of the sign of the interaction $U$, despite the presence of $\mathcal N=4$ excitations in the system.
    }
    \label{fig5:std-transport-weak}
\end{figure*}

\section{Source-to-drain transport}
\label{sec:sourcetodrain}

In this section we address the configuration shown in Fig.~\ref{fig1:schematic}(b), where the system Hamiltonian is $\mathcal{\hat H} = \mathcal{\hat H}_\text{BH} + \mathcal{\hat H}_\text{SD}$ [cf. Eqs~\eqref{eqn:hamiltonianBH} and \eqref{eqn:H_SD}]. We investigate the dynamics of the average number operator in the source $n_S(t)=\langle \hat n_S \rangle=\braket{\hat a^\dagger_S\hat a_S}$, chain $\langle \hat N \rangle = \sum_{i=1}^M \langle \hat n_i \rangle$, and in the drain $n_D(t)=\langle \hat n_D \rangle=\braket{\hat a^\dagger_D\hat a_D}$. Here, we discuss the case where the source and drain are weakly coupled to the chain ($J'=0.1J$). The system initially has all excitations in the source: $|\psi_0\rangle = (\hat{a}^\dagger_S)^{\mathcal N}  |0\rangle/\sqrt{\mathcal{N}!} $, with $|0\rangle$ being the vacuum state (we remind that $\mathcal N$ denotes the total number of excitations in the system, including source and drain).

\subsection{Transmons resonant to the resonator}
\label{sec:sourcetodrainResonant}

We begin by discussing the case where the frequencies of the source and drain resonators match the qubit frequency in the transmon array. Figures~\ref{fig5:std-transport-weak}(a)-(c) display the time evolution of the expectation value of the number operator in the source, chain, and drain for various values of on-site interaction $U$, and a system with an even number of sites in the chain ($M = 4$). For most values of $|U|$, the excitations coherently oscillate between the source and the drain with a period almost insensitive to the sign and the strength of the interaction $U$. In particular, the number of excitations in the chain remains typically quite low, $\langle \hat N \rangle\ll 1$, throughout the evolution; in other words, the transport through the chain is fast compared to the typical source depletion, due to the weak coupling $J'\ll J$. (For some specific values of $|U|$, the dynamics is more complex; this is explored in more detail in \cref{app:stdspurious}.)

Figures~\ref{fig5:std-transport-weak}(g)-(i) show that, for an odd number of sites, the dynamics is rather different: coherent oscillations dominated by a single frequency can only be seen for $U=0$, while the time evolution at finite interaction is more complex. In comparison with the case of even sites [\cref{fig5:std-transport-weak}(a)-(c)], there are more frequency components in the oscillations, and faster exchange of excitations between the resonators and the transmon array. Moreover, even at zero interaction, a few excitations are populating the transmon array. 

The qualitative differences between the even and odd case for the parity in the number of sites can be understood by analyzing the non-interacting case ($U=0$). In this case, the Bose-Hubbard Hamiltonian reduces to the tight-binding model, and can be diagonalized by moving to momentum space, \textit{i.e.}, $\hat{H}_{BH}(U=0)= \sum_{k = 1}^{M} \epsilon_k \hat b^\dagger_k \hat b_{k}$, see \cref{app:TBdiag}. Here, the operators $\hat b^\dagger_k$ and $\hat b_k$ respectively creates and annihilates a boson of energy $\epsilon_k=\w_{01}+2J\cos[\pi k/(M+1)]$, with $k=1,\dots,M$. 
The difference between the parities stems from the absence (presence) of a single-particle eigenstate in the chain resonant with the source and drain levels in the even (odd) case. Indeed, the condition $\epsilon_k=\w_{01}$ requires $k=(M+1)/2$, which can be satisfied only for odd values of $M$, since $k$ is an integer number. This resonance is responsible for the faster dynamics observed in the odd case. We note that this parity effect has been discussed for a single excitation traveling along a spin-chain for purposes of perfect state transfer~\cite{wojcik2005unmodulated}. 

\begin{figure*}
    
    \centering
    \includegraphics[width=\textwidth]{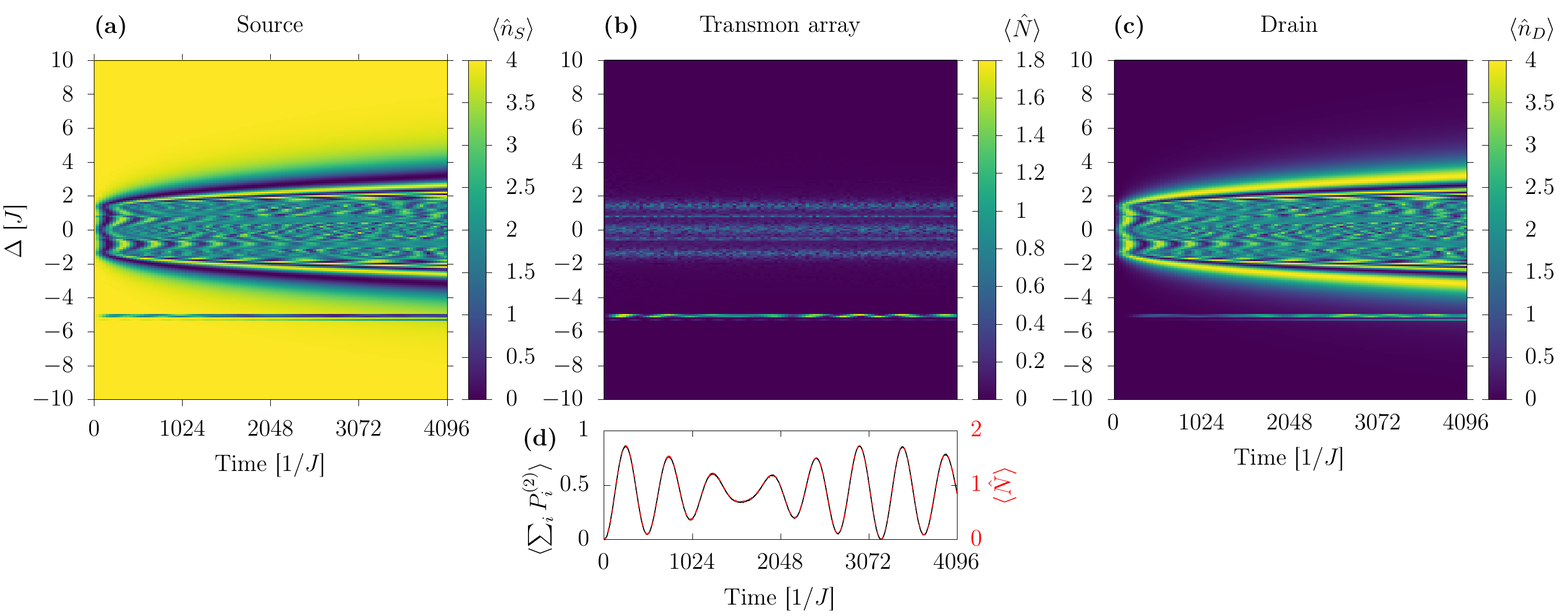}
    \caption{Source-to-drain photon transport vs source/drain and transmon qubits detuning. Long-time evolution of the occupation numbers of the (a) source, (b) chain, and (c) drain for an attractive ($U=-10J$) on-site interaction. The source and drain are weakly coupled to the main system, with $J' = 0.1J$. A level of strong resonances can be seen for $\Delta=\w_r-\w_{10} \approx U/2$, when two-particle states are accessed within the central chain. 
    (d) Fidelity of the instantaneous state with $2$-particle states for $\Delta=-5J$ (solid black). The evolution of the occupation number $N$ in the chain (dashed red) follows closely the dynamics of the fidelity, showing how $2$-particle effects are dominant in the transport.  Parameters: odd-parity chain with $M=3$ sites, and $\mathcal{N}=4$ particles.}
    \label{fig6:std-transport-sweep-mu-attractive}
\end{figure*}

More precisely, for $U=0$ and odd-parity, the dynamics for small chain sizes can be approximated by exclusively accounting for the transport mediated by the resonant level. Within this description, the expectation values of the density in the source and drain are approximately expressed as (see Appendix~\ref{app:odd-chain})
\begin{align}
    \frac{n_S(t)}{\mathcal{N}} =& \cos^4 \left(\frac{J'}{\sqrt{M+1}} t\right),
    \label{eqn:density-source-U0}\\
    \frac{n_D(t)}{\mathcal{N}} =& \sin^4 \left(\frac{J'}{\sqrt{M+1}} t\right),
    \label{eqn:density-drain-U0}
\end{align}
while the density in the chain can be obtained through particle conservation $N(t)=\mathcal{N}-n_S(t)-n_D(t)=\mathcal{N}\sin^2\left(2J't/\sqrt{M+1}\right)/2$. These expressions give quite an accurate description of the dynamics for $M=3$, as displayed in the zero-interaction cuts of the density plots in Figs.~\ref{fig5:std-transport-weak}(j)-(l).
Clearly the above expressions do not describe the dynamics in the case of finite interaction $U\neq 0$, see Figs.~\ref{fig5:std-transport-weak}(g)-(i).

For even chains, the evolution at zero interaction is more involved, including more energy levels. In first approximation, the dynamics is the result of beatings between a slow frequency $\w_-$ and amplitude $\sim\mathcal N$, modulating a fast oscillation with frequency $\w_+$ and smaller amplitude (see Appendix~\ref{app:even-chain}):
\begin{align}
    \frac{n_S(t)}{\mathcal{N}} =& \left[\frac{1+\alpha}{2}\cos(\w_- t)+\frac{1-\alpha}{2}\cos(\w_+ t)\right]^2,
    \label{eqn:density-source-U0-even}\\
    \frac{n_D(t)}{\mathcal{N}} =& 
    \left[\frac{1+\alpha}{2}\sin(\w_- t)-\frac{1-\alpha}{2}\sin(\w_+ t)\right]^2,
    \label{eqn:density-drain-U0-even}
\end{align}
with $\alpha=(\w_+-\w_-)/(\w_++\w_-)\approx 1$. In this approximation, $N(t)=\mathcal{N}-n_S(t)-n_D(t)=\mathcal{N}(1-\alpha^2)\,\sin^2[(\w_++\w_-)t/2]\ll \mathcal{N}$. Including only the two single-particle energy levels closest to the frequency of the resonator, we obtain $\w_\pm=J \sin\left(\frac{\pi}{2(M+1)}\right)(\sqrt{1+2\beta^2}\pm 1)$, with $\beta=(J'/J)\,\text{cot}[\pi/2(M+1)]\sqrt{2/(M+1)}$. The zero-interaction approximation (black triangles) and the evolution obtained through exact diagonalization (solid) are shown in Figs.~\ref{fig5:std-transport-weak}(d)-(f). The analytical expression overestimates the slower oscillation frequency $\w_-$. However, upon including the energy shift due to level repulsion from the adjacent energy levels (see Appendix~\ref{app:even-chain}), we obtain the more accurate expression for the slow frequency $\w_-=J'^2/(J[1+MJ'^2/(2J^2)])$ [orange points in Figs.~\ref{fig5:std-transport-weak}(d)-(f)].

\subsection{Resonator and qubit frequency detuning}
Here, we explore the source-to-drain dynamics when the resonators, with frequency $\w_r$, are detuned from the transmons in the chain, with energy separation $\w_{01}$. In the density plots of Figs.~\ref{fig6:std-transport-sweep-mu-attractive}(a)-(c), we explore the dynamics for different values of the detuning $\Delta=\w_r-\w_{01}$, keeping fixed the interaction $U<0$ and the number of excitations $\mathcal N$. We identify different regimes: for $|\Delta|<2J$, the evolution of the excitation number in the resonators (and the excitations in the transmon array) depends non-monotonically on $|\Delta|$. Source-to-drain transport occurs readily here, with a periodicity modulated by the detuning.
For larger values, \textit{i.e.}, $|\Delta|\geq 2J$, the excitations are not transferred from the source to the drain on the considered timescale. An exception is the specific detuning $\Delta \approx -5 J$, which is also characterized by a more prominent occupation of the chain. 

These features can be understood in terms of the analytical analysis of the resonant case. In particular, by changing the detuning, the single-photon energy resonates with the single-particle energy in the chain for $\w_r=\w_{01}+2J\cos [\pi k/(M+1)]$, for which $|\Delta|< 2J $. Outside this region, the transport becomes typically extremely small since the large detuning from the single-particle levels makes the time evolution progressively slower. 

The behavior at $\Delta \approx -5 J$ denotes a resonance between the two-particle states in the chain and the Fock states in the resonator. Indeed, for the source and drain to be resonant with states which are a superposition of states characterized by 
$N$ particles in a specific site, energy conservation imposes (we disregard the tunnelling energy in this reasoning)
 \begin{equation}
 \w_r \mathcal{N}\approx \w_r (\mathcal{N}-N)+\w_{01}N+\frac{U}{2} N \left(N - 1\right),    
 \end{equation}
with $\w_r \mathcal{N}$ being the energy of the Fock state prepared in the source at $t=0$. This relation immediately implies $\Delta = U(N-1)/2$; for the parameters used in Fig.~\ref{fig6:std-transport-sweep-mu-attractive}, this gives $\Delta=-5J$ for the two-particle states. Such a resonance has been very recently experimentally observed by~\citet{fedorov2021photon}.
We remark that our considerations are valid for values of $U$ such that the ground-state band is well separated from the extended states, see Fig.~\ref{fig1:schematic}(a) 
(for instance, $U \lesssim -2J$ for $N = 4$~\cite{naldesi2019rise}). Resonances with higher number particle states $N>2$ are, in principle, possible. While these resonances are difficult to identify at weak coupling $J'/J \ll 1$, we have found evidence for them in simulations at stronger coupling (not shown here).

To confirm the multi-particle nature of the resonance observed at $\Delta=-5J$, we compute the expectation value of the projector onto the two-particle subspace $P^{(N=2)}$ (\textit{i.e.}, the fidelity of the time-evolved state with any state where there are only two particles on any one transmon and an arbitrary number in the resonators), shown in Fig.~\ref{fig6:std-transport-sweep-mu-attractive}(d) (black solid curve). This quantity reproduces exactly the dynamics of the excitations number in the chain [red dashed curve in Fig.~\ref{fig6:std-transport-sweep-mu-attractive}(d)]; this feature suggests that for $\Delta=-5J$, the source to drain transport is mediated by two-excitation states. 

\section{Experimental considerations}
\label{app:decoherence}

In Secs.~\ref{sec:model} and \ref{sec:Dynamics}, we discussed the spectral properties of the Bose-Hubbard model describing the array chain and, consequently, the dynamics of localized excitations. We highlighted the stability features of pinned solitons compared to boson stacks. In this section, we comment on typical values of model parameters and possible sources of non-idealities in realistic experimental setups. First, we note that the pinned soliton should be prepared on a timescale shorter than the typical tunneling time $1/J$. For coupling frequency of the order $J/{2\pi}\sim 1\ldots100$~MHz~\cite{hacohen2015cooling,ma2019A,fedorov2021photon,Gong2021experimental,karamlou2022quantum}, the soliton should then be prepared in less than $1~\mu$s, which is achievable with state-of-the-art optimized pulses. In this respect, the preparation of the pinned soliton may be assisted with engineered dissipation, recently exploited for the realization of a Mott insulator state of photons~\cite{ma2019A}. Being the typical on-site interaction in the range of $U/2\pi= -150\ldots -300$~MHz~\cite{hacohen2015cooling,fedorov2021photon,Blok2021quantum},  $|U|/J$ can  assume values to explore both the intermediate ($|U|\lesssim |U_C|$)  and the strong-interaction ($|U|\gg |U_C|$) regimes discussed in this article. Finally, we comment that the soliton pinning can be achieved, for instance, with flux-tunable control of the split-transmon frequency, with typical modulations of the order of $1$~GHz~\cite{RevModPhys.93.025005}.

We remark that our closed-system modeling with fixed excitation number $N$ is valid on a timescale where relaxation and excitation effects can be disregarded. Firstly, excited states can decay due to relaxation phenomena; in particular, $T_1^{N\to N-1}$ denotes the timescale on which a transmon prepared in the level with $N$ plasmons loses an excitation. Secondly, dephasing effects can degrade the interference pattern in the evolution of the quantum soliton state; the pure dephasing time $T_{\varphi}^{(N)}$ determines the coherence time of the $N$-plasmons state $T_{2}^{N\to N-1}$ in combination with the relaxation rate $1/T_{2}^{N\to N-1}=1/T_{\varphi}^{(N)}+1/2T_{1}^{N\to N-1}$. In a single transmon coupled to a three-dimensional cavity~\cite{Paik2011observation}, these multi-excitation states can live up to a few tens of $\mu$s, with reported values as high as $T_1^{2\to 1}\approx 40~\mu$s, $T_1^{3\to 2}\approx 30~\mu$s, $T_1^{4\to 3}\approx 20~\mu$s
~\cite{peterer2015coherence}, and with typical coherence times of $T_2^{2\to 1}\approx 30~\mu$s, $T_2^{3\to 2}\approx 10~\mu$s, $T_2^{4\to 3}\approx 2~\mu$s. In two-dimensional implementations of transmon arrays, similar values  were recently reported for the second excited state of the qubit, with the lifetime $T_1^{2\to 1}=30-40~\mu$s, and coherence time $T_2^{2\to 1}=10-20~\mu$s~\cite{Blok2021quantum} (coherence times become as large as $T_2^{2\to 1}\approx 70~\mu$s with spin-echo~\cite{krantz2019quantum}). Considering a state-of-the-art transmon array and previous characterization of single qubits, two-particle solitons are achievable with current implementations. Because of the slow-down properties discussed in Sec.~\ref{sec:velocity}, at present addressing solitons with higher excitation-number dynamics may be more challenging. 

Additional limitations are associated with the requirement of identical transmon frequencies, as assumed in the dynamics of multi-particle states considered in Sec.~\ref{sec:Dynamics}. The frequency of transmon qubit $i$ can be approximately expressed as $\omega_{01}^{(i)}\approx\sqrt{8E_J^{(i)}E_C^{(i)}}-E_C^{(i)}$, an approximation valid if the Josephson energy is much larger than capacitive energy, $E_J^{(i)}\gg E_C^{(i)}$. During fabrication, the capacitive energy can be controlled with a high degree of accuracy, being associated with several micrometers to fractions of millimeter size structures; the Josephson energy, on the other hand, depends on the properties of a nanometer-thick oxide barrier over a tunnel junction area typically smaller than a micron square, so its value is less reproducible. The potential frequency mismatch, effectively determining a disordered array, can be addressed by using split-transmons, where the junction is replaced by a SQUID and the qubit frequency is controlled on-chip via an out-of-plane magnetic field~\cite{krantz2019quantum}. When tuned to the same frequency, the transmons would generally be no longer operating at their ``sweet-spot'' in flux-tunability (integer multiples of $\Phi_0/2$ for an asymmetric split transmon~\cite{hutchings2017Tunable}, where $d\omega_{01}(\Phi)/d\Phi=0$), making the system more sensitive to magnetic flux noise~\cite{ithier2005Decoherence}. In this situation, the decoherence in the system is dominated by pure dephasing, resulting in $T_2<T_1$~\cite{hutchings2017Tunable}; for this reason, below we neglect excitation/decay processes and discuss the dynamics with keeping fixed the total number of particles. 
Recent theoretical work~\cite{Busel2023Dissipation} supports this choice, as the analysis performed there also shows that dephasing is the dominant decoherence channel.  

To simulate the effect of dephasing on the dynamics of excitations, we proceed as follows.
Since the spectral density of the flux noise is peaked at low-frequency, we perform the simulations with a quasistatic
noise; in particular, we perform ensemble averages over several time evolutions. For each trajectory, the transmon frequencies $\w_{01}^i$ are randomly sampled from a Gaussian distribution with mean $\omega_{01}$, and standard deviation inversely proportional to the pure dephasing time $\sigma_\omega=\sqrt{2}/T_\varphi$~\cite{ithier2005Decoherence}. The change in magnetic flux affects the Josephson energies $E_J^{(i)}$ and, thus, the frequencies $\omega_{01}^{(i)}$ of the transmons, while the capacitive energy (that is, the interaction strength $U$) is uniform along the chain. The coupling between the transmons also becomes site-dependent, $J\to J_{i,i+1}=J\sqrt{\omega_{01}^{(i)}\omega_{01}^{(i+1)}}/\omega_{01}$ (the scaling factor originates from rewriting the capacitive coupling between charges in terms of creation/annihilation operators). As in the plots displayed in Figs.~\ref{fig2:bosonvsstack}(d),~2(e), and~\ref{fig4:ramp-heights}(c), we consider a system of $N=3$ particles, $U=-3J$, and $M=19$ sites. With a tunneling coupling of $J=20\ldots 100$~MHz and considering the dephasing times reported in Ref.~\cite{hutchings2017Tunable} for tunable transmons ($T_\varphi \approx5\ldots15~\mu$s, with longer coherence for less tunable qubits), the standard deviation is in the range $\sigma_\omega\sim 0.001J\ldots 0.014J$. In the numerical simulations presented below, we explore up to the worst-case scenarios of $\sigma_\w=0.05J$ for expansion and $\sigma_\w=0.1J$ for directional transport.

\begin{figure}
    \centering
    \includegraphics[width=\linewidth]{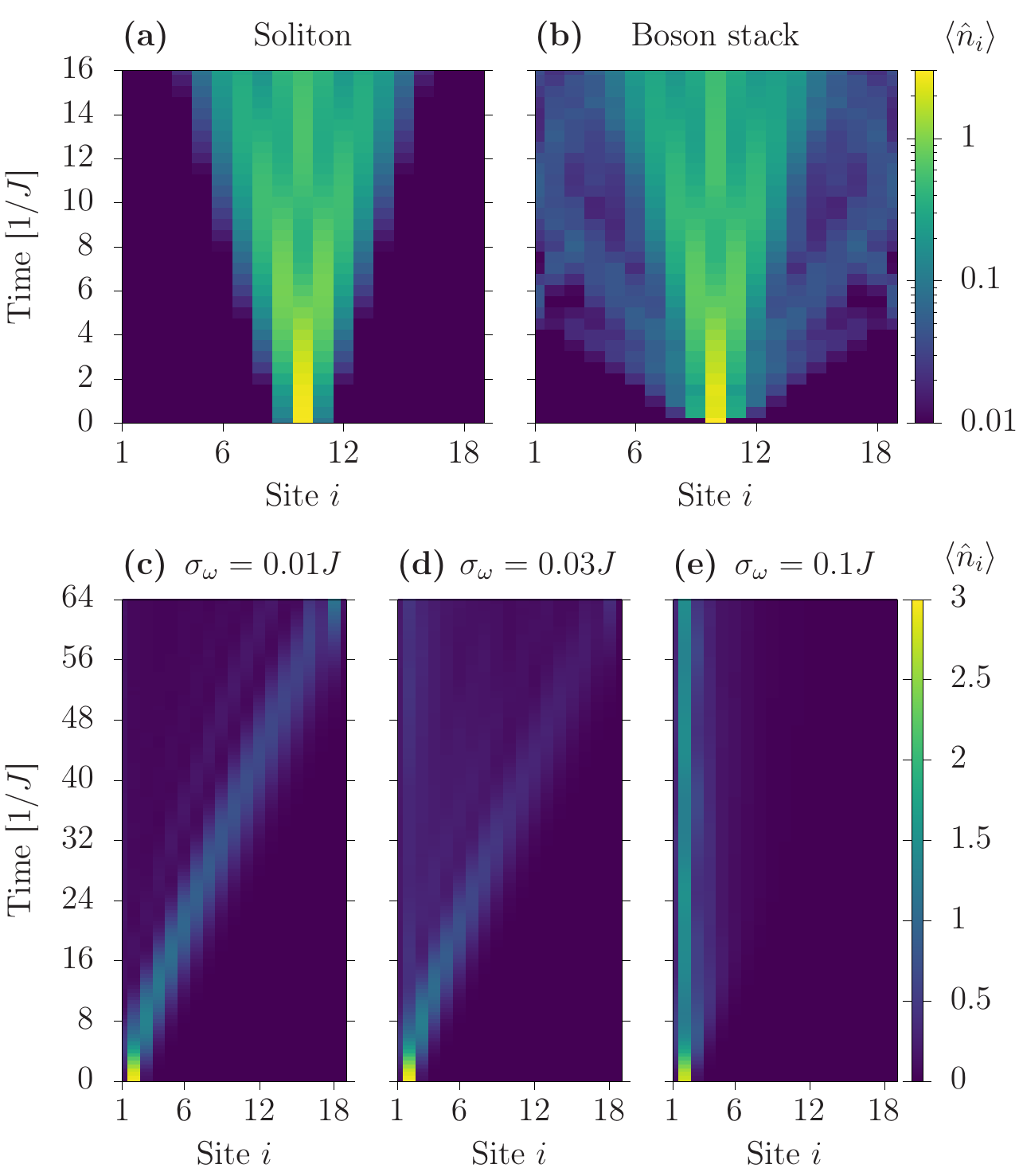}
    \caption{
    Density evolution with time of (a) a quantum soliton and (b) a boson stack prepared in the center of the system with $\sigma_\w=0.05J$ determining the pure dephasing rate. In both cases, the interference pattern observed in \cref{fig2:bosonvsstack}(d)-(e) has blurred. The previously-observed differences between a quantum soliton and boson stack still hold under decoherence. (c)-(d) The evolution of a quantum soliton prepared in the second site from the left [similar to \cref{fig4:ramp-heights}(c)], with different values of pure dephasing rate $\sigma_\w/\sqrt{2}$. With large decoherence, a localization effect is observed. Parameters are the same as in \cref{fig2:bosonvsstack}(d)-(e) and \cref{fig4:ramp-heights}(c) to aid comparison. The soliton is prepared with $\mu_{i_\text{pin}} = \w_{01}^i-\mu_\text{pin}$, with $\mu_\text{pin}\gg\sigma_\w$. The inter-site tunneling $J$ has been scaled as a result of the changes in Josephson energy of each of the transmons -- see text.
    }
    \label{fig:dephasing}
\end{figure}

Figures~\ref{fig:dephasing}(a)~and~(b) show the time evolution of a quantum soliton and a boson stack, respectively, averaged over 100 repetitions~\footnote{Numerical simulations show that, with this range of parameters, averaging over 100 realizations is sufficient to sample evenly from the probability distribution. We also performed preliminary simulations with a single qubit to show that 100 repetitions are sufficient to observe Ramsey decay.} with $\sigma_\omega=0.05J$. The bosonic excitations are prepared in the center of the chain, and the soliton is pinned with $\mu_{i_\text{pin}} = \w_{01}^{(i)}-\mu_\text{pin}$; the pinning procedure is robust against the frequency fluctuations, being  $\mu_\text{pin}\gg\sigma_\omega$. Comparing with Figs.~\ref{fig2:bosonvsstack}(d)~and~(e), we see that the differences between the quantum soliton and boson stack expansions continue to hold under relatively strong dephasing: as in the absence of dephasing, the density profile spreads outwards from the center of the chain with velocity $v\sim \tilde J$ in both cases, but the boson stack continues to exhibit fast-moving ($v\sim J$) single-particle dynamics. The dephasing affects both the boson stack and quantum soliton dynamics in similar ways: the sharp interference pattern displayed in Figs.~\ref{fig2:bosonvsstack}(d)~and~(e) in the density evolution fades with increasing $\sigma_\omega$, and completely disappears for $\sigma_\omega\gtrsim 0.1J$ (not shown here). 

Next, we explore the effect of dephasing on the directional motion of the quantum soliton. Figures~\ref{fig:dephasing}(c)-(e) show the dynamics of a quantum soliton initially pinned at site $i=2$ for different values of $\sigma_\omega$. In the absence of decoherence, the soliton moves towards the right end of the chain with no substantial expansion of the soliton wavepacket [cf. Fig.~\ref{fig4:ramp-heights}(c)]. For $\sigma_\omega\lesssim 0.01J$, as in typical experimental conditions, the directional dynamics is not qualitatively affected by dephasing. For completeness, we also explore larger dephasing rates; in particular, we observe localization of the excitations for large values of $\sigma_\omega$ [Fig.~\ref{fig:dephasing}(e)]. This phenomenon resembles the localization of excitations under strong disorder $D \gtrsim 2\tilde{J}/N$~\cite{Mansikkamaki2021Phases}, with $D$ the disorder strength (which plays the role of $\sigma_\omega$ in our notation). We numerically verified that the localization occurs for any arbitrary choice of pinning site (not shown). We conclude this section with a few final observations. For dephasing rates such that $\sigma_\omega\sim\mu_\text{pin}$ (not shown), the soliton pinning in a specific site $i_\text{pin}$ is no longer possible: indeed, in many instances, the excitations would localize in site $i$, whereby $\w_{01}^{(i)}<\w_{01}^{(i_\text{pin})}-\mu_\text{pin}$. Similar considerations apply to the case of the ramp protocol discussed in \cref{sec:directional}; in particular, the ramp is unaffected by dephasing for $\mu_\text{ramp} \gg \sigma_\w$ (as is the case for optimal ramp slope).

\section{Conclusions}
\label{sec:conclusions}

In this work, we considered the dynamics of bosonic excitations (plasmons) in an array of capacitively coupled transmons. The system's dynamics is governed by a suitable Bose-Hubbard model describing correlated bosons with attractive interaction.

We analyzed two different dynamical protocols. In the first, we considered  excitations initially localized around a specific site in the transmon chain. The required pinning energy $\mu_\text{pin}$ is realized by tuning the frequency of a single transmon. In this way, we engineer a lattice bright soliton; such a state differs from a stack of bosons localized in a single site~\cite{SilveriBeyondHardcore}. Indeed, the pinned soliton dynamically evolves in the form of a superposition of solitons localized at different sites. Therefore,  we find that the soliton, as such, is dynamically stable. This property arises because the solitonic bound states are protected by a characteristic energy gap in the Bose-Hubbard spectrum which increases with the interaction $|U|$. 
Boson stacks, instead, are found to be stable only for sufficiently large interaction.
The characteristic soliton velocity is found to be a universal function of a specific combination of interaction strength and number of particles for any finite $U$. This feature provides a remarkable extension of the scaling found for boson stacks~\cite{Mansikkamaki2021Phases} and reflects the aforementioned bright soliton stability for any finite attractive interaction.

In the second protocol, we prepared the source resonator in the $N$-photon Fock state and transferred excitations to the transmon array by means of a weak capacitive coupling.
In the case in which the qubits are resonant with the resonator (the case of odd length of the chain of transmons), we observe transport with almost no dependence on the sign of $U$; all transport occurs through the resonant zero-energy state which is independent of interaction. Indeed, the solitonic states cannot be accessed from this zero-energy state. Yet, by suitably detuning transmon and resonator frequencies, states containing multiple particles can, in principle, be accessed. However, due to the weak single-particle tunneling between the resonators and the chain, the dynamics in the chain is comparably fast, and only few-particle states in the chain are relevant in the dynamics. The strength of interaction determines the degree of detuning required to reach these interaction-dependent states.

Finally, we note that the low-atom-number regime is challenging to achieve in cold atoms and atomtronics settings \cite{amico2022colloquium}. 
As discussed in this work, though, a low excitation number is a  natural condition in superconducting circuits implementations. Therefore, our results can provide a first step in the identification of currently-unobserved phenomena, such as the fractional flux quantization predicted to occur in ring-shaped condensates \cite{naldesi2022enhancing}.

\acknowledgments

The authors would like to thank Rainer Dumke for fruitful discussions.

\appendix

\section{Spectral decomposition of soliton versus stack}\label{app:state-tomography}

In Sec.~\ref{sec:stackvspinned}, we discussed the dynamics of a localized quantum soliton compared to a boson stack. In particular, we showed that the quantum soliton expands into a superposition of multi-boson states while the boson stack displays single-particle features associated with scattering (or extended) states. Here, we remark on the differences between a quantum soliton and a boson stack by inspecting the spectral decomposition of the two states. We consider the expansion of the states on the basis given by the eigenstates $\ket{\psi_n}$ of the Bose-Hubbard model \cref{eqn:hamiltonianBH} (and no pinning, $\mu_i=\w_{01}$), \textit{i.e.}, $\ket{\phi}=\sum_n \braket{\psi_n|\phi}\ket{\psi_n}$. 

\begin{figure}[!b]
    \centering
    \includegraphics[width=\linewidth]{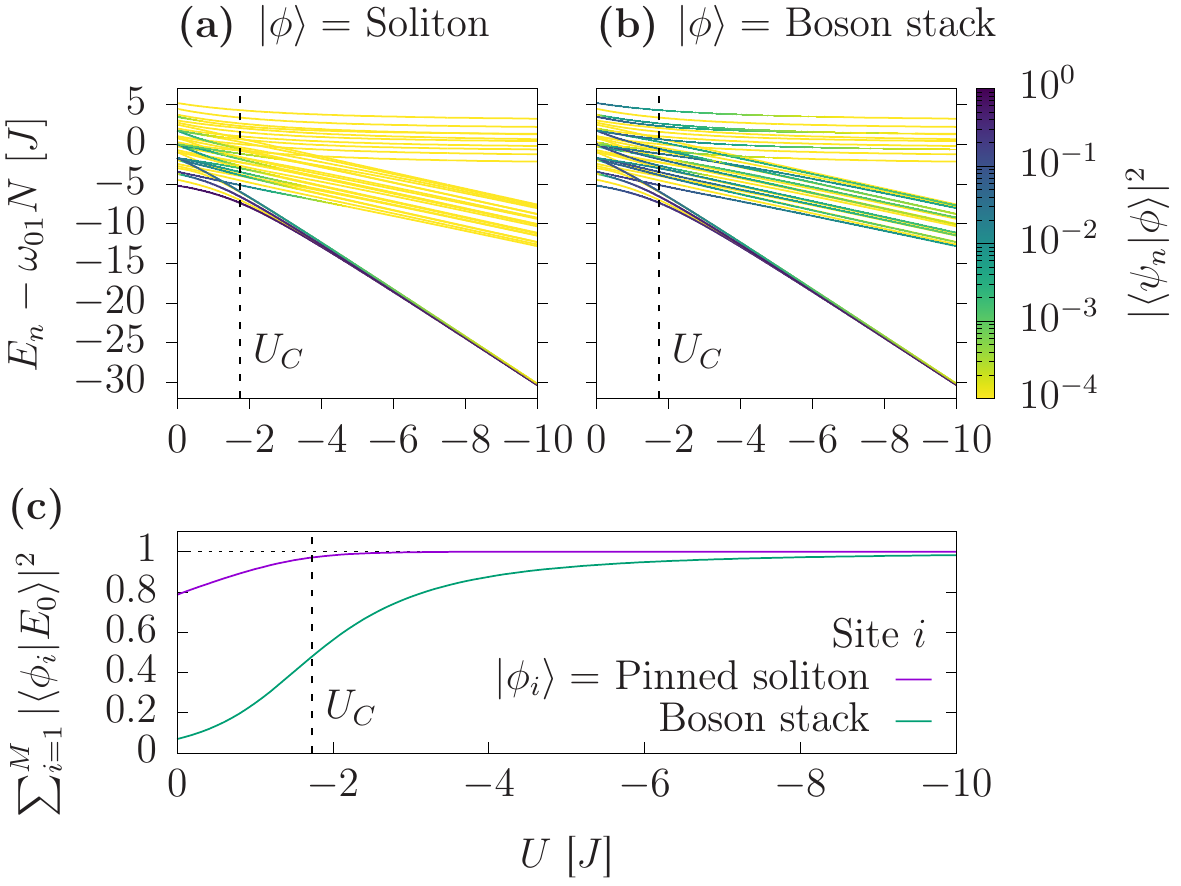}
    \caption{Spectral decomposition of a quantum soliton and a boson stack. (a)-(b) Spectrum of the Bose-Hubbard Hamiltonian \cref{eqn:hamiltonianBH} (with no pinning, $\mu_i=\w_{01}$) for $N = 3$ particles and $M = 5$ sites versus interaction. The coloration of each eigenvalue curve is associated with the weight of the spectral decomposition $|\langle \psi_n|\phi\rangle|^2$, where $|\phi\rangle$ is (a) a quantum soliton [the ground state of \cref{eqn:hamiltonianBH} with $\mu_{i}=\w_{01}-\mu_\text{pin}\delta_{i,(M+1)/2}$] in the center of the system or (b) a boson stack $\left(\hat b^\dagger_{(M+1)/2}\right)^N|0\rangle$ on the central site. For values of $|U|>|U_C|$ where the band gap has formed, the soliton almost exclusively occupies bound eigenstates (cf. \cref{fig1:schematic}). In contrast, the boson stack continues to occupy scattering states even at larger $|U|$. 
    To enhance visibility, eigenvalue curves with higher weight are plotted on top of those with lower weight and the lowest value in the colorbar is used for all the states with  $|\langle\psi_n|\phi\rangle|^2\leq 10^{-4}$.
    (c) Fidelity of the ground state of the unpinned Bose-Hubbard Hamiltonian \cref{eqn:hamiltonianBH} (for $N=3, M=15$) with a superposition of (purple) pinned quantum solitons and (green) boson stacks across all sites. The quantum soliton is pinned with the usual method of $\mu_{i_\text{pin}}=\w_{01}-\mu_\text{pin}$. For $|U|>|U_C|$, the ground state of the Bose-Hubbard model consists primarily of a superposition of pinned solitons. 
    }
    \label{fig:app-state-tomography}
\end{figure}

In \cref{fig:app-state-tomography}, we display the spectrum of the Bose-Hubbard model \cref{eqn:hamiltonianBH} for $N=3$ particles and $M=5$ sites as a function of the interaction $U$. The coloration of each curve reflects the overlap $|\braket{\psi_n|\phi}|^2$ of the initially prepared state $\ket{\phi}$ and the eigenstate corresponding to each eigenvalue (see the color bar on the right side of the figure). We compare a localized soliton [panel (a)] in the center of the chain, \textit{i.e.}, the ground-state of the pinned Hamiltonian [\cref{eqn:hamiltonianBH} with $\mu_{i}=\w_{01}-\mu_\text{pin}\delta_{i,(M+1)/2}$], with a boson stack [panel (b)] $\ket{\phi}=\ket{N_{(M+1)/2}}=\left(\hat b^\dagger_{(M+1)/2}\right)^N|0\rangle$. We note that, while the quantum soliton (almost) exclusively occupies the localized-states band -- especially for $|U|>|U_C|$ -- the boson stack's overlap with the scattering states is still significant even for moderate to large interactions $|U|/J$. The distinct spectral decompositions of the boson stack and the soliton are reflected in the different dynamics shown in \cref{fig2:bosonvsstack}(d)-(e): the scattering components of the boson stack expand more rapidly compared to the soliton.

\Cref{fig:app-state-tomography}(c) shows the expectation values of the projector $P=\sum_{j=1}^M\ket{\phi_j}\bra{\phi_j}$ onto i) the subspace generated by pinned solitons, \textit{i.e.}, $\ket{\phi_j}$ is the ground state of~\cref{eqn:hamiltonianBH} with $\mu_i=\omega_{01}-\mu_{\rm pin}\delta_{i,j}$ (purple line); and ii) the subspace generated by boson stacks $\ket{\phi_j}=\ket{N_j}$ (green line), computed on the ground state $\ket{E_0}$ of the unpinned Bose-Hubbard model \cref{eqn:hamiltonianBH}. The expectation value for the pinned soliton states is substantially larger than for boson stacks when $|U|<|U_C|$ and close to one for $|U|>|U_C|$ (small deviation from unity is due to the pinning perturbation); in other words, for $|U|> |U_C|$ the ground state of the system is a superposition of quantum solitons but not of boson stacks. Only in the limit of strong interaction $|U|/J\gg 1$ does the expectation value for the boson stacks also approach unity; indeed, in this regime, the pinned soliton width is much smaller than the lattice size ($R^2\ll 1$), making the soliton closer to a boson stack. 

\section{Width of a pinned quantum soliton in the strong pinning limit}\label{app:soliton-width-derivation}

Here we derive an approximate expression for the width of the pinned soliton $R^2(0)$ in the limit of strong pinning $\mu_{\rm pin}\gg J$. The chemical potential profile reads $\mu_i=\w_{01}-\mu_{\rm pin}\delta_{i,i_{\rm pin}}$, hence we can replace the last term in the Hamiltonian Eq.~\eqref{eqn:hamiltonianBH} with $-\mu_{\rm pin}\hat{n}_{i_{pin}}$ subtracting the constant term $\w_{01}N$. In this respect, the pinning acts as a local disorder, which is known to localize the excitations in the strong disordered limit~\cite{Mansikkamaki2021Phases}. In particular, the excitation density in the chain sites decays exponentially with $|i-i_{\rm pin}|$. In first approximation, we can assume that the pinning and the nearest-neighbor sites have finite excitation-density. That is, we adapt the treatment given by \citet{Mansikkamaki2021Phases} as first-order perturbation theory for the localized phase. We use the notations $\ket{N_{i}} = \left(b^\dagger_i\right)^{N_i} \ket{0}/\sqrt{N_i!}$ and $\ket{N_{i},N_{j}} = \left(\hat{b}^\dagger_i\right)^{N_i}\left(\hat{b}^\dagger_j\right)^{N_j} \ket{0}/\sqrt{N_i!N_j!}$ to denote the state with $N_i$ (and $N_j$) excitations at site $i$ (and $j$) and no excitation elsewhere. Then we write the state of the pinned soliton as:
\begin{equation}\label{eq:approxQS}
    \ket{\psi_{\rm sol}} = \beta \left[|N_{i_{\rm pin}}\rangle + |\psi_{i_\text{pin}+1}\rangle + |\psi_{i_\text{pin}-1}\rangle \right]
\end{equation}
where $\beta$ is a normalization factor and
    \begin{equation}
    |\psi_{i_\text{pin}\pm 1}\rangle = - \frac{J\sqrt{N}}{\mu_\text{pin} +|U|(N-1)}|(N-1)_{i_\text{pin}}, 1_{i_\text{pin}\pm 1}\rangle.
    \label{eqn:NNdensity}
    \end{equation}
As discussed in the main text, the width of a pinned soliton is given by
    \[
    \sqrt{R^2(0)} = \sqrt{\frac{1}{N}\sum_{i=1}^M \langle \hat n_i\rangle (i-i_\text{pin})^2},
    \]
Upon taking the expectation value
in the approximate state of Eq.~(\ref{eq:approxQS}), $\braket{\hat n_i}\approx\braket{\psi_{\rm sol}|\hat n_i|\psi_{\rm sol}}$, we find immediately    
 \begin{align*}
        \sqrt{ R^2(0)} = \frac{\sqrt{2} J}{\mu_\text{pin} + |U|\ (N-1)}
        \label{eqn:widthlargemu}
\end{align*}

\section{Determination of the asymptotic expansion velocity}\label{appendix:steady-state}

\begin{figure}
    \centering
    \includegraphics[width=\linewidth]{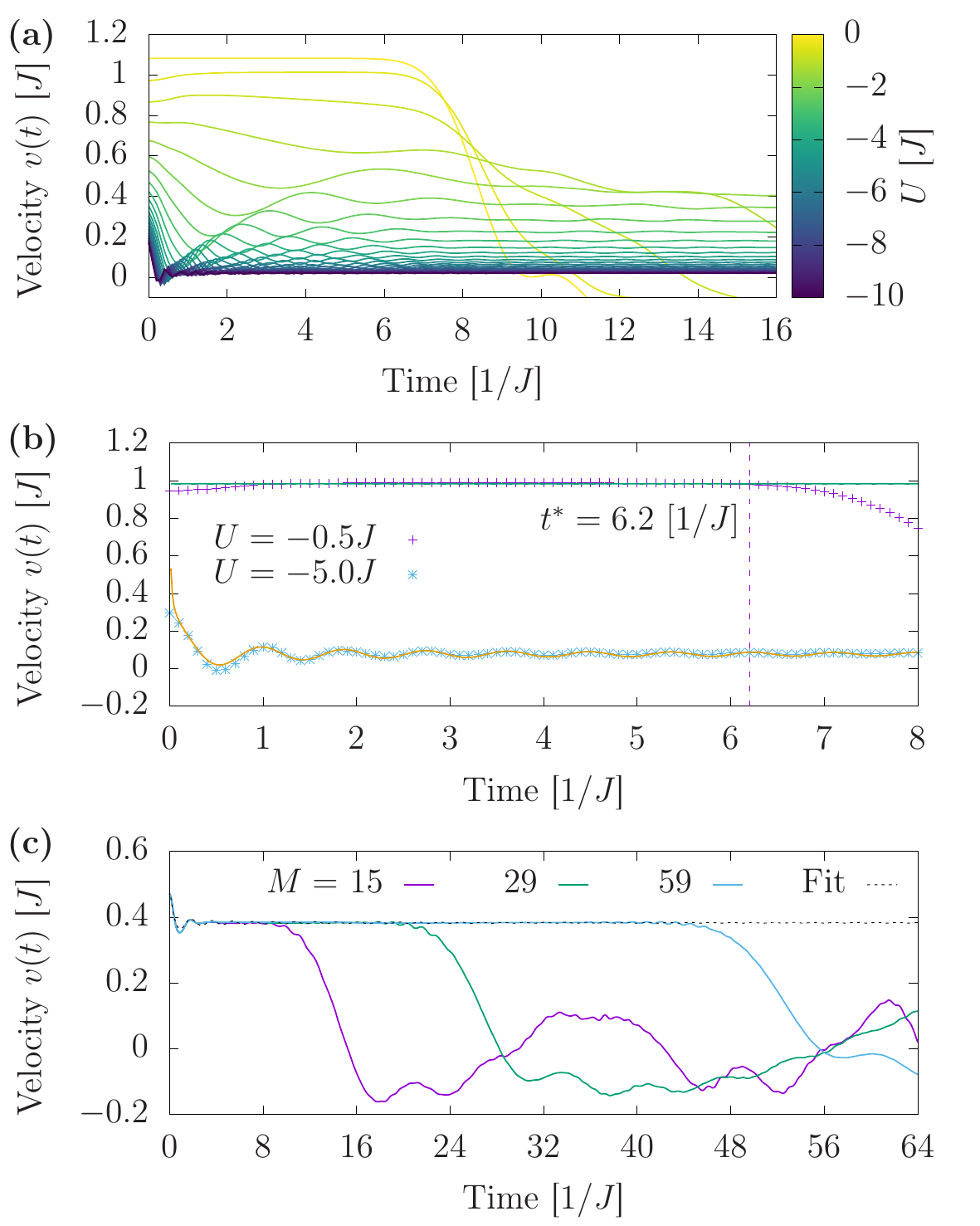}
    \caption{Comparison of the RMSD velocity \cref{eqn:velocity} for $N=3$ and \textbf{(a)} values of $U$ in the range $-10J\le U\le 0$, \textbf{(b)} chosen values of $U$ plotted alongside the fitted function \cref{eqn:fitting-velocity}. The dashed line shows the approximate time $t^*$ when the boundary effects are first seen in the $U=-0.5J$ data series ($t^*=6.2/ J$). \textbf{(c)} The RMSD velocity for $N = 2,\, U = -6J$ for different values of chain length $M$. The dashed line shows the fit of the  $M=29$ curve by using \cref{eqn:fitting-velocity}, according to the procedure outlined in the text [$t^*=6.2/J$ as in panel (a)]; clearly the fitting procedure (and so the steady state velocity $v_\infty$) is robust to variation in chain length, provided an appropriate maximum fitting time $t^*$ is chosen before the velocity decays.
    }
    \label{fig:velocity-fitting-samples}
\end{figure}

In Sec.~\ref{sec:velocity} we investigated the expansion velocity Eq.~\eqref{eqn:velocity} as a function of the (attractive) interaction $U$. As mentioned in the main text, this velocity is a function of time, while the quantity displayed in \cref{fig3:velocity} depends on the interaction only. Here we briefly discuss how the plots in~\cref{fig3:velocity} are obtained, starting from the simulation of the Bose-Hubbard Hamiltonian within the exact-diagonalization technique.
First, we numerically compute, with a finite difference scheme for the time derivative, the time-dependent velocity as expressed by Eq.~\eqref{eqn:velocity}; the time evolution of the velocity is displayed in \cref{fig:velocity-fitting-samples}(a). For $U=0$, the expansion velocity is constant for $t<t^*$, while for $t>t^*$ this quantity is reduced due to the reflection by the boundaries of the chain. In other words, $t^*$ represents the maximum time up to which finite-size effects can be disregarded in the expansion of the localized excitations. We remark that we aim to simulate the expansion velocity in an infinite-size chain and we are here limited due to computational resources. For the case of $U=0$, we can simply average the velocity for time smaller than $t^*$, \textit{i.e.}, $\bar v=1/t^*\sum_j v(t_j)\delta t$, where $\delta t$ is the time discretization step in the calculation. For finite interaction, evolution is generally more complicated. First, the time at which the finite-size effects are visible is a function of the interaction $t^*=t^*(U)$, reflecting the dependence of the expansion velocity on $U$. We choose $t^*\approx 6.2/J$ by inspecting the time evolution at small $|U|<|U_C|$, see Fig.~\ref{fig:velocity-fitting-samples}(b). This value is also consistent with the ballistic expansion taking time $M/4J$, where $M=29$ for the plots in Figs.~\ref{fig:velocity-fitting-samples}(a)-(b). Moreover, for $|U|\gtrsim |U_C|$ and $t<t^*(U)$, there is a transient time where the velocity oscillates significantly around a value $v_\infty$, representing the steady state value of the expansion velocity for an infinite-size chain.

\begin{figure}
    \centering
    \includegraphics[width=\linewidth]{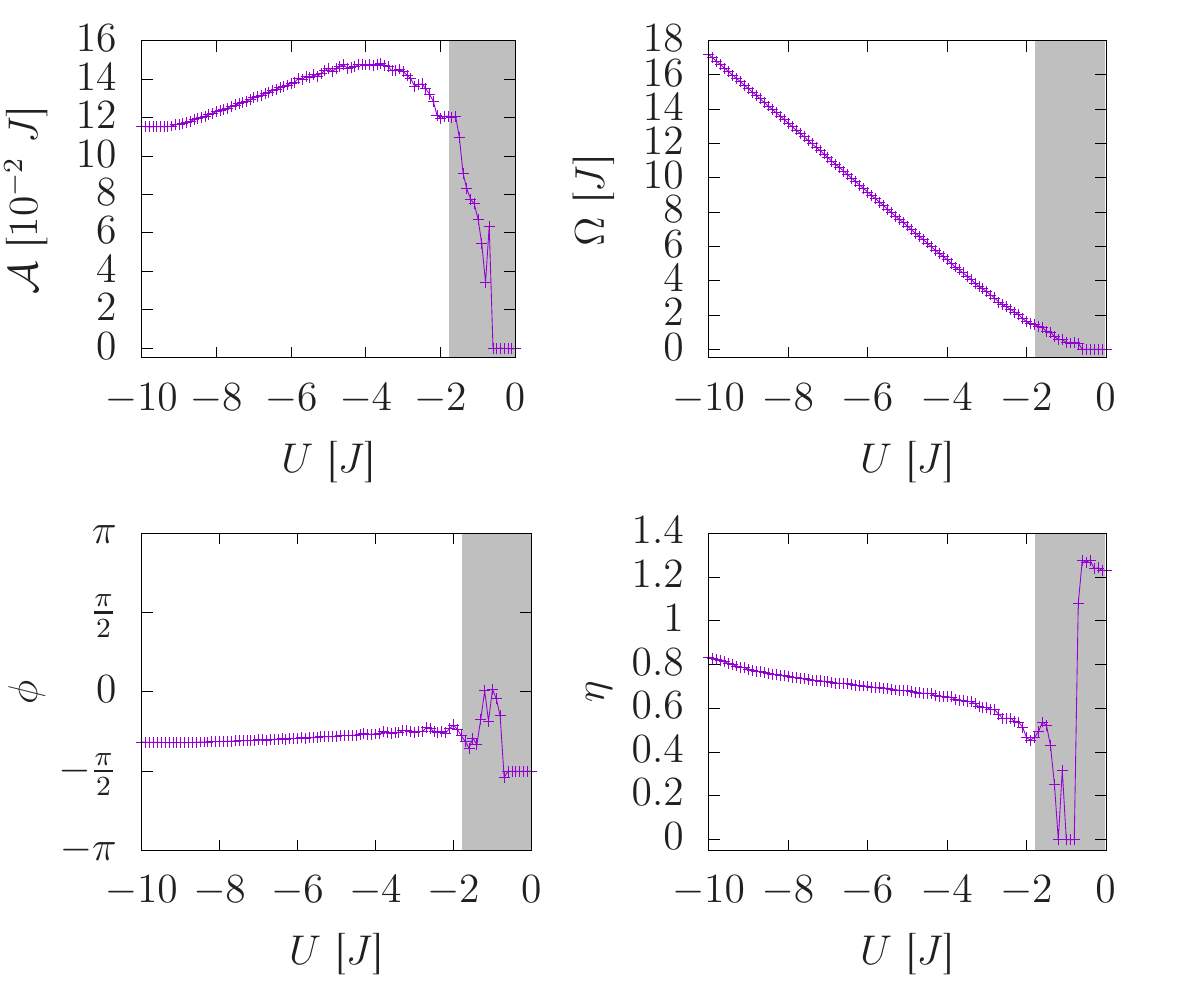}
    \caption{Fitting parameters of the RMSD velocity fitting function \cref{eqn:fitting-velocity} for an expanding quantum soliton consisting of $N=3$ particles. 
    The shaded areas identify the regions with no gap in the spectrum $|U|<|U_C|$. There, the amplitude $\mathcal{A}$ decreases quickly towards zero and therefore the remaining parameters $\phi,\,\Omega$, and $\eta$ become indeterminate.}
    \label{fig:velocity-fitting}
\end{figure}

To account for these short-time transients, we performed a nonlinear fitting of the RMSD velocity \cref{eqn:velocity} as a function of time to the 5-parameter equation~\cite{Boschi2014bound,naldesi2019rise}
\begin{equation}
v(t) = v_\infty + \mathcal{A}\cos(\Omega\,t + \phi) / (\Omega\,t)^\eta
\label{eqn:fitting-velocity}
\end{equation}
within the aforementioned steady-state window $t_\varepsilon < t < t^*$ (where $t_\varepsilon$ is a short initial time period to discount the larger initial peak in velocity due to settling from quench of the pinning) selected by inspection of both the density dynamics and $v(t)$ in order to extract the long-time steady-state velocity $v_\infty$. 
Figure~\ref{fig:velocity-fitting-samples}(c) shows the time evolution of the RMSD velocity, \cref{eqn:velocity}, for $N=2,\, U=-6J$, and different chain lengths $M$. Up to a cut-off time $t^*$ that depends on the chain length, the velocity curves coincide. Hence, the extracted steady-state velocity $v_\infty$ of Fig.~\ref{fig3:velocity} is independent of the chain length, as shown by the fit of the $M=29$ curve [dashed line in Fig.~\ref{fig:velocity-fitting-samples}(c)] obtained using \cref{eqn:fitting-velocity}. In other words, since the velocity is not affected by the chain length up to the time $t^*$ that grows with $M$. The steady-state velocity $v_\infty$ can be accurately estimated even using the shortest considered length $M=15$.

\Cref{fig:velocity-fitting} shows the evolution of the fitting parameters in \cref{eqn:fitting-velocity}, where $\mathcal{A}$ is the amplitude, $\Omega$ is the frequency, $\phi$ is the phase (modulo $2\pi$), and $\eta$ is the damping of the fitted transient oscillations.
We identify two regimes, depending on the strength of $|U|$. The first (with low on-site interaction and where there is no band-gap in the spectrum [as introduced in Figs.~\ref{fig1:schematic}(a) and~\ref{fig3:velocity}(a)]) leads to the expansion velocity $v(t)$ being less well-defined due to the ready occupation of scattering states. This is seen as a lack of initial stability and near-ballistic dynamics in the expanding quantum soliton, where the amplitude $\mathcal{A}$ of the fitted transient oscillations is near-zero. Thus, the fitting function is insensitive to the phase, frequency, and damping of the oscillations and their respective fitted values are indeterminate (shown in the shaded regions of \cref{fig:velocity-fitting}). It is important to note that, in this region, the asymptotic expansion velocity $v_\infty$ is still well-defined with little error, with the fitting tending towards a time-average of this parameter [see the $U=-0.5J$ series in \cref{fig:velocity-fitting-samples}(b)]. In the second regime, with the increase of $|U|$ and higher retention of bound states due to the presence of a band-gap in the spectrum [as per \cref{fig3:velocity}(b)], the aforementioned transients in $v(t)$ start to become apparent [as seen in \cref{fig:velocity-fitting-samples}(a) and exemplified in the $U=-5J$ series of \cref{fig:velocity-fitting-samples}(b)]. This is evident from the finite value of the oscillation's amplitude $\mathcal{A}$ with respect to $U$, along with the frequency $\Omega$ of the oscillations increasing linearly with $-U$. The damping $\eta$ of these transient oscillations increases slightly with $-U$, while the phase $\phi$ remains roughly constant.

\section{Critical interaction for band gap formation}\label{app:UC}

In Sec.~\ref{sec:velocity}, we discussed how the propagation of the localised excitation displays features of universality in terms of a certain combination of $U/J$ and $N$, \textit{i.e.}, $|U/J|^{N-1}(N-1)!/N$. In particular, for different particle number $N$, the critical value for the opening of the gap appears approximately constant $\sim 2-3$, as visible in the inset of \cref{fig3:velocity}(c). This feature suggests using the scaling for an effective single-particle boson stack from Ref.~\cite{SilveriBeyondHardcore} as a way to estimate the critical interaction $U_C$ for an arbitrary particle number.

\begin{figure}
    \centering
    \includegraphics{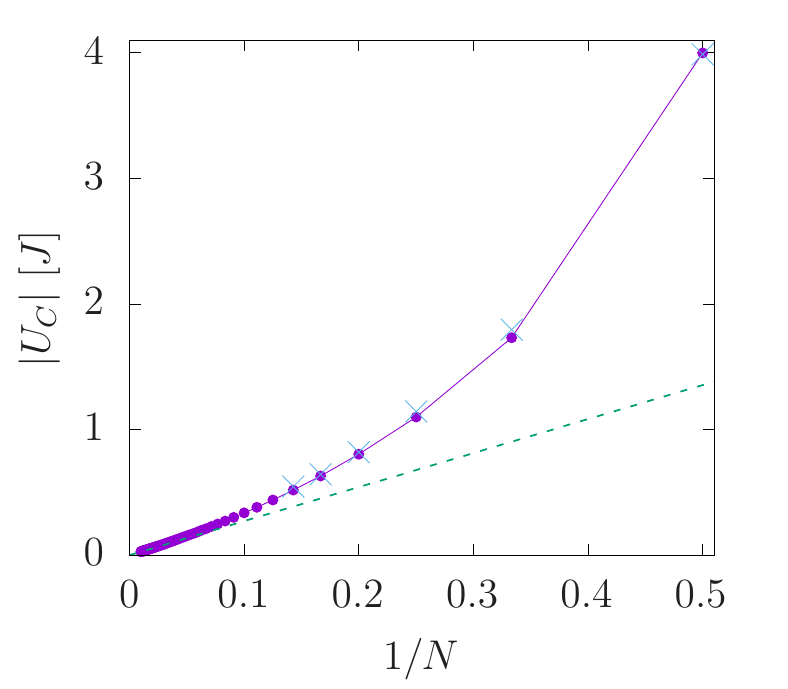}
    \caption{Critical value of the interaction $U$ for the formation of the band-gap in the Bose-Hubbard spectrum, as a function of particle number. The connected data points showing $|U_C(N)|$ are given by \cref{eqn:UcVsN}. The dashed green line shows scaling for large $N$, \textit{i.e.}, $|U_C| = eJ/N$. The light blue numerical crosses are extracted from Ref.~\cite{naldesi2019rise}.}
    \label{fig:u_c}
\end{figure}

More precisely, we consider the following expression for the critical interaction,
\begin{equation}
|U_C(N)| = J\left[\frac{\alpha N}{(N-1)!}\right]^\frac{1}{N-1}.
\label{eqn:UcVsN}
\end{equation}
Above, we fix the constant
$\alpha = 2$ to match the value of the critical interaction $|U_C(N=2)|=4J$ for periodic lattices, where an exact solution is available~\cite{Boschi2014bound}. Notably, the analytical formula Eq.~\eqref{eqn:UcVsN} reproduces with a good degree of accuracy the values obtained numerically in Ref.~\cite{naldesi2019rise} (see Appendix of the reference) for every particle number. In the limit of large particle number $N\gg 1$, the critical interaction asymptotically scales as $|U_C| \approx eJ/N$ (see dashed line in \cref{fig:u_c}), with $e\approx 2.718\dots$ being Euler's number.

\section{Source-to-drain dynamics for selected values of $|U|$}\label{app:stdspurious}

In this Appendix, we explore the interaction effect in the source-to-drain configuration discussed in Sec.~\ref{sec:sourcetodrain}. Specifically, we consider chains of even length and investigate the dynamics at those values of $|U|$ for which there is a non-negligible occupation of the transmon array, such as $|U|/J\simeq 3.5$ and $|U|/J\simeq 2.5$, see \cref{fig5:std-transport-weak}(b). 

\begin{figure}
    \centering
    \includegraphics[width=\linewidth]{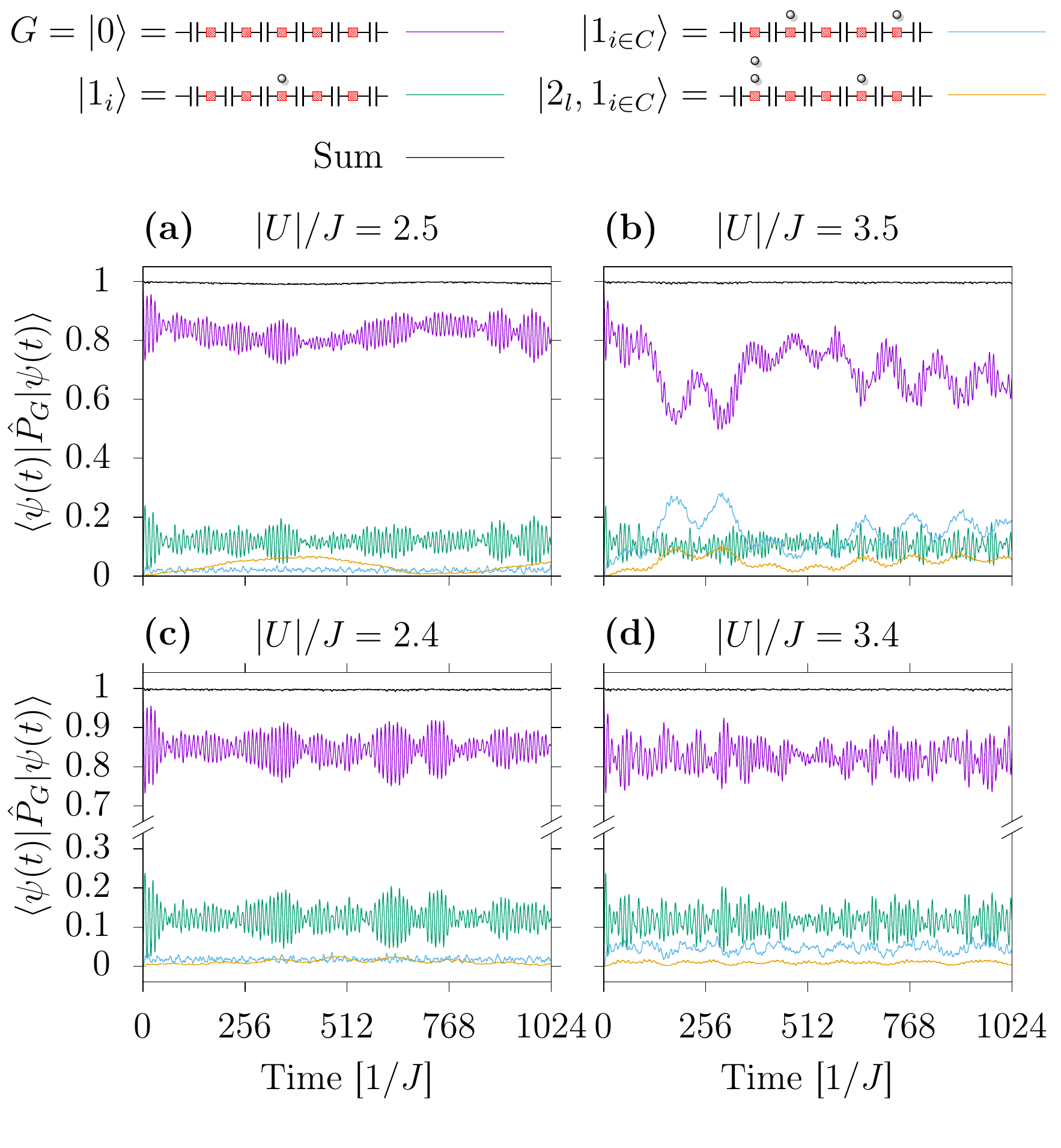}
    \caption{Time evolution of the projection $\hat{P}_G$ over groups of states $G$ detailed in the main text for $\mathcal N=4$ particles and $M=4$ sites in the chain. At most values of $|U|$, exemplified by the cases $|U|/J=2.4$ and $|U|/J=3.4$ in panels (c) and (d), respectively, the dynamics is largely determined by states where there is no or one particle in the chain. At certain values of $|U|$, such as (a) $|U|/J=2.5$ and (b) $|U|/J=3.5$, the contribution of states with higher particle number (in particular, those with two particles in one site) cannot be neglected.
    }
    \label{fig:appdxprojectionspurious}
\end{figure}

To characterize the role of interaction, in Fig.~\ref{fig:appdxprojectionspurious} we display the time evolution of the states involved in the dynamics at $|U|/J=2.5$ [panel (a)] and $|U|/J=3.5$ [panel (b)]. More specifically, we consider the amplitude $\braket{\psi(t)|\hat{P}_G|\psi(t)}$ of the time-evolved state $\ket{\psi(t)}$ projected onto specific subspaces $G$: i) empty transmon array (purple), generated by the states $\ket{\phi}$ with $\hat{b}_i\ket{\phi}=0$, ii) single particle in the transmon array, generated by states of the type $\ket{n_S}\otimes\ket{1_i}\otimes\ket{n_D}$, with $|1_i\rangle = \hat{b}^\dagger_i|0\rangle$ and $n_S+n_D=\mathcal N-1$, iii) two or more single particles in the array (but no sites with higher occupation), generated by states $\ket{n_S}\otimes\ket{1_{i\in C}}\otimes\ket{n_D}$ where $C\subseteq\{1,\ldots,M\}$ is a set of sites in the chain, $\hat{b}^2_j|1_{i\in C}\rangle = 0$ for any $j$, and $n_S+n_D<\mathcal{N}-1$, 
and iv) a single two-particle stack plus additional individual excitations in the array, spanned by states $\ket{n_S}\otimes\ket{2_l,1_{i\in C}}\otimes\ket{n_D}$, with $\ket{2_l,1_{i\in C}}$ denoting a state with two-bosons in site $l$ and at least a particle in site $i\neq l$, such that $n_S+n_D<\mathcal{N}-2$.

We observe that with high probability the chain is either empty (purple) or singly-occupied (green) during the evolution. However, multi-particle effects -- exemplified by two bosons in a single transmon plus additional single excitations -- can contribute significantly to the dynamics, having at some times a larger probability than single occupation states for $|U|/J=3.5$; thus, they determine the noticeable occupation displayed in Fig.~\ref{fig5:std-transport-weak}(b). 
From the formation of these multi-particle states, one can see that the system undergoes a complex dynamics, where particles do not solely traverse the chain (source-to-drain) independently but, effectively, can also hop over each other, leading to in-phase time oscillations between states in the chain of the type $\ket{1_{i\in C}}$ and $\ket{2_l,1_{i\in C}}$ evident in Fig.~\ref{fig:appdxprojectionspurious}(b). Note that even though the four groups of states considered do not generate the entire Hilbert space, the sum of the four probabilities is close to unity at all times; therefore, additional states do not contribute significantly to the dynamics. For comparison, in Fig.~\ref{fig:appdxprojectionspurious}(c)-(d), we consider the time evolution of the projected state for slightly smaller interaction values, respectively $|U|/J=2.4$ and $|U|/J=3.4$, where the density occupation is small [cf. \cref{fig5:std-transport-weak}(b)]. Notably, we observe that the multi-particle effects are significantly suppressed. 

\begin{figure}
    \centering
    \includegraphics[width=\linewidth]{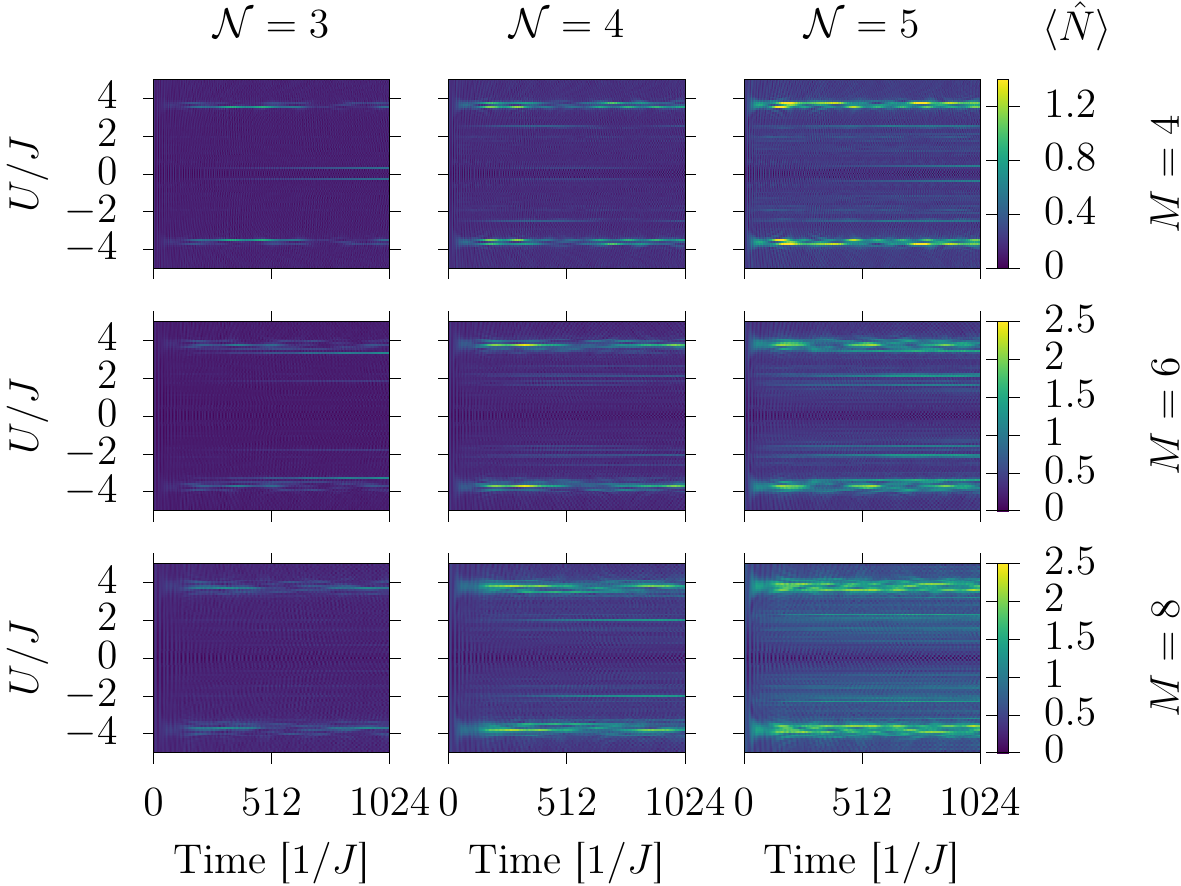}
    \caption{
    Time evolution of the total occupation number in the chain for different values of $\mathcal N$ and $M$. Generally, when increasing the number of particles $\mathcal N$ or sites $M$, more values of $U$ display relatively large density in the chain.
    }
    \label{fig:appdxdensitychaingrid}
\end{figure}

We next explore how the features associated with multiple particles in the chain are modified by the size of the chain $M$ and the total particle number $\mathcal{N}$, see \cref{fig:appdxdensitychaingrid}. Overall, both increasing $\mathcal N$ and $M$ enhance these features.  Indeed, by increasing $\mathcal N$, a larger number of particles can stack within the chain, so the multi-particle states are more likely to be occupied. Also, longer chains support additional oscillating modes, and these additional modes likely increase the number of possible routes for multi-particle states to form by tunneling, leading to more values of $|U|$ displaying relatively high density in the chain. Notably, the aforementioned features at $|U|/J=3.5$ and $|U|/J=2.5$ are robust and well-identifiable for all $\mathcal{N}$ and $M$, despite the appearance of additional resonances for larger $\mathcal{N}$ and $M$. The precise characterization of the full $U$ dependence is quite complex and goes beyond the scope of this work.

\section{Diagonalization of the tight-binding Hamiltonian in a finite chain.}\label{app:TBdiag}

For zero interaction $U = 0$ and zero disorder $\mu_i=\w_{01}$ in the transmon array, the Hamiltonian
    Eq.~\eqref{eqn:hamiltonianBH} for a fixed number of excitations $N$ reduces to the tight-binding Hamiltonian
    \begin{equation}
        \hat H_{\rm TB}= \hat H_{BH}(U=0) -\w_{01}N =J \sum_{i = 1}^{M - 1} (\hat b^\dagger_{i+1} \hat b_{i} + \hat b^\dagger_{i} \hat b_{i+1}).   
    \label{eqn:tight-binding}
    \end{equation}
The tight binding Hamiltonian can be standardly diagonalized in the momentum space, with a specific Fourier-like transformation. For completeness, here we show the diagonalization for the case of interest, following the derivation in Ref.~\cite{Zawadzki2017Symmetries}.
    We start expressing the site-dependent second-quantization operators in terms of momentum space operators, 
    \begin{equation}
     \hat b^\dagger_i = \frac{\sqrt{2}}{\sqrt{M+1}} \sum_{k = 1}^{M} \sin\left(\frac{\pi k i}{M+1}\right) \hat b^\dagger_k.
     \label{eqn:momentumOperators}
    \end{equation}
Note that the specific transformation accounts for the finite size of the chain, as expressed by the conditions $\hat b^\dagger_0 = \hat b^\dagger_{M+1} = 0$. Inserting the expressions Eq.~\eqref{eqn:momentumOperators} into Eq.~\eqref{eqn:tight-binding}, we obtain
\begin{align}
    & \hat{H}_{\rm TB} = \label{eqn:tight-binding_momentum} \\
    & \frac{2J}{M+1}\! \sum^{M-1}_{i=1} \sum^M_{k, k'=1} \!\!\left[ \sin\left(\frac{\pi k (i+1)}{M+1}\right) \sin\left(\frac{\pi k' i}{M+1}\right) \right] \hat b^\dagger_k \hat b_{k'}  \nonumber\\
    &+\! \frac{2J}{M+1} \! \sum^{M}_{j=2} \sum^M_{k, k'=1} \!\!\left[ \sin\left(\frac{\pi k (j-1)}{M+1}\right) \sin\left(\frac{\pi k' j}{M+1}\right) \right] \hat b^\dagger_k \hat b_{k'} \nonumber    
\end{align}
where $j = i + 1$ in the second sum. Note that the lower limit of the $j$ sum can be changed to $j = 1$, and the upper limit of the $i$ sum can be set to $M$ since the added terms are zero. 
Relabeling $j\to i$ in the second term in Eq.~\eqref{eqn:tight-binding_momentum}, and using the sine addition formula $\sin(x+y) + \sin(x-y) = 2\sin(x)\cos(y)$ with $x=\pi k i/(M+1)$, $y=\pi k/(M+1)$, the tight-binding Hamiltonian is diagonalized   
\begin{equation}
    \mathcal{\hat H}_{\rm TB} = 2J \sum_{k = 1}^{M} \cos\left(\frac{\pi k}{M+1}\right) \hat b^\dagger_k \hat b_{k}.
    \label{eq:HTBDiag}
\end{equation}
Indeed the double-sum over $k,k'$ disappears due to the discrete orthogonality relation of the trigonometric sine functions
\begin{align*}
        \sum_{i=1}^{M} \sin\left(\frac{\pi k i}{M+1}\right) \sin\left(\frac{\pi k' i}{M+1}\right) &= \,\frac{M+1}{2}\delta_{k, k'};
\end{align*}
with $k, k' = 1, \ldots, M$.

\section{Source-drain number occupation dynamics at zero-interaction}
\label{app:odd-chain}

For the determination of the source-to-drain dynamics, we mainly follow and generalize the treatment given in the supplemental material of Ref.~\cite{PhysRevA.100.041601}.
Since we are interested in treating the non-interacting case $U=0$, 
we start by rewriting the source and drain Hamiltonian of Eq.~\eqref{eqn:H_SD}, in terms of momentum operators in the transmon chain, which diagonalize the tight-binding Hamiltonian as detailed in Appendix~\ref{app:TBdiag},
\begin{align}
\mathcal{\hat H}_{\text{SD}} & = \w_r (\hat{a}_S^\dagger\hat{a}_S+\hat{a}_D^\dagger\hat{a}_D) +J' \sqrt{\frac{2}{M+1}} \\
&  \times\sum_{k=1}^{M} \sin\left(\frac{\pi k}{M+1}\right) \left( \hat a_S \hat b_k^\dagger - (-1)^k \hat a_D \hat b_k^\dagger + \text{h.c.}\right).\nonumber
\end{align} 
We assumed resonant source and drain resonators $\w_S=\w_D=\w_r$. 

For $U=0$ and the specific initial state considered in the main text, \textit{i.e.}, all excitations are initially in the source $\ket{\psi_0}=\ket{\mathcal{N}_S}$, a further simplification applies. Indeed, both the Hamiltonian and the initial state can be written as a $\mathcal{N}$-tensor-product of single-excitation states. Hence, in deriving the excitation dynamics, we can consider the case of a single excitation $\mathcal N=1$. The expressions obtained can be extended to the case $\mathcal N>1$ by multiplying them by $\mathcal{N}$. For $\mathcal N=1$, it is convenient to perform the calculation in the basis $\{\ket{S},\ket{k},\ket{D}\}$, where the state $\ket{k}$ corresponds to having the excitation in the chain with momentum $k=1,\dots,M$, while $\ket{S} (\ket{D})$ denote the state with the excitation is in the source (drain); in this notation, our initial condition is $\ket{\psi_0}=\ket{S}$. 

\subsection{Resonant case}
We start by discussing the case where one of the momentum eigenstates in the transmon chain is resonant with the resonators, \textit{i.e.}, $\epsilon_{\bar k}=\w_r$.
In first approximation, one can study the dynamics in
a reduced subspace where only the resonant state is included to account for the chain. This leaves us working in the three-dimensional subspace generated by the states $\{\ket{S}, \ket{\bar{k}}, \ket{D}\}$. The truncated Hamiltonian reads 
\[
\mathcal{H_F} = \w_r \mathbb{I}_3+\alpha_{\bar k}
  \begin{pmatrix}
    0 & 1 & 0\\
    1 & 0 & 1\\
    0 & 1 & 0
  \end{pmatrix},
\]

where $\mathbb{I}_3$ is the $3\times 3$ identity matrix and
\begin{equation}
\alpha_{\bar k} = J' \sin\left(\frac{\pi \bar{k}}{M+1}\right) \sqrt{\frac{2}{M+1}}.
\end{equation} 
Our initial state is $\ket{\Psi(0)} = \ket{S} = \{1,0,0\}$ and its time evolution is

\[ \ket{\Psi(t)} = \sum_j e^{-\imath E_j t} \ket{\Psi_j} \langle\Psi_j |S\rangle. \]

One can then compute the expectation value of the number operators on the time-evolved state:
\begin{align}
\langle \Psi(t) | \hat n_S | \Psi(t)\rangle &= \cos^4\left(
\frac{\alpha_{\bar k}}{\sqrt{2}}t
\right); \label{eq:resns} \\
\langle \Psi(t) | \hat n_D | \Psi(t)\rangle &= \sin^4\left(
\frac{\alpha_{\bar k}}{\sqrt{2}}t
\right)
; \label{eq:resnd} \\
\langle \Psi(t) | \hat N | \Psi(t)\rangle &= \frac{1}{2} \sin^2\left(
\frac{2\alpha_{\bar k}}{\sqrt{2}}t 
\right)
\label{eqn:resonantDensities}.
\end{align}

These results can be specialized to the case of zero detuning $\Delta=\w_r-\w_{01}=0$. In this case, there is a resonant level for chains with an odd number of sites, $M+1$ mod 2 =0, with $\bar{k} = \frac{M+1}{2}$, and the expressions in Eqs.~\eqref{eq:resns}-\eqref{eq:resnd}
can be used with $\alpha_{\bar{k}}= J'\sqrt{2/(M+1)}$, giving Eqs.~\eqref{eqn:density-source-U0}-\eqref{eqn:density-drain-U0} in the main text (once multiplied by $\mathcal{N}$).

\subsection{Off-resonant case.}
\label{app:even-chain}

Here we discuss the dynamics of the occupation number when there is no single-particle energy level in the chain resonant with the frequency of the resonators. In this case, we only focus on the case of zero detuning $\Delta=0$ and even number of sites in the chain, $M\ \text{mod}\ 2 = 0$. Generalizing the method discussed in the resonant case, we approximate the dynamics considering only the two states of the chain closer in energy to the resonators' frequency which, in the present case, are the one with momentum $k_1=\frac{M}{2}$ and $k_2=\frac{M}{2}+1$. The Hamiltonian in the subspace spanned by the states $\{\ket{S}, \ket{k_1}, \ket{k_2}, \ket{D} \}$ reads:

\begin{equation}
\mathcal{H_F} = \w_r \mathbb{I}_4+
  J \sin\left[\frac{\pi}{2(M+1)}\right]\begin{pmatrix}
    0 & \beta & \beta & 0\\
    \beta & 2 & 0 & -\beta c\\
    \beta & 0 & -2 & \beta c\\
    0 & -\beta c & \beta c & 0
  \end{pmatrix},
  \label{eqn:truncated4x4}
\end{equation}
where 
\begin{equation} \label{eq:betadef}
    \beta = \frac{J'}{J} \text{cot}\left(\frac{\pi}{2(M+1)}\right) \sqrt{\frac{2}{M+1}},
\end{equation}
$c = (-1)^{M/2}$, and $\mathbb{I}_4$ is the $4\times4$ identity matrix.
It can be easily proved that the eigenvalues of Eq.~\eqref{eqn:truncated4x4} are symmetric about $\w_r$, \textit{i.e.}, $\{\w_r-\w_+,\w_r-\w_-,\w_r+\w_-,\w_r+\w_+\}$ with
\begin{align}
    \w_{\pm}&= J\sin\left[\frac{\pi}{2(M+1)}\right](\sqrt{2\beta^2 + 1} \pm 1)
    \label{eqn:even-chain-w-pm}
\end{align}
Computing the time evolution in this subspace, the densities in the source and drain are expressed by Eqs.~\eqref{eqn:density-source-U0-even}-\eqref{eqn:density-drain-U0-even} of the main text.

The approach described above turns out to be inaccurate in the determination of the slow-frequency $\w_-$ for sizes $M>2$, since the level repulsion from states outside the considered subspace has non-negligible effects. We obtain a more accurate determination of $\w_-$ following a method adapted from the appendix of \citet{PhysRevA.100.041601}.

For $\mathcal{N}=1$,  the state of our system is generally expressed as $\ket{\Psi} = \alpha_S\ket{S} + \sum_{k=1}^M \alpha_k \ket{k} + \alpha_D \ket{D}$. Starting from the time-independent Schr\"odinger equation $\mathcal{\hat H}\ket{\Psi} = (\w_r+E)\ket{\Psi}$ and solving for the coefficients $\alpha_k$, we obtain coupled equations for $\alpha_S$ and $\alpha_D$:
\begin{align}
    E \alpha_S &= \frac{J'^2}{J(M+1)} \sum_{k=1}^M \frac{\sin^2\left(\frac{\pi k}{M+1}\right)\left(\alpha_S - (-1)^k \alpha_D\right)}{\frac{E}{2 J} - \cos\left(\frac{\pi k}{M+1}\right)};
    \label{eqn:coefficientSource}
    \\
    E \alpha_D &= -\frac{J'^2}{J(M+1)} \sum_{k=1}^M \frac{\sin^2\left(\frac{\pi k}{M+1}\right)\left((-1)^k \alpha_S - \alpha_D\right)}{\frac{E}{2 J} - \cos\left(\frac{\pi k}{M+1}\right)}.
    \label{eqn:coefficientDrain}
\end{align}
Since we are interested in a low-energy solution $E=\w_-\ll 2J\sin(\pi/(M+1))$, we consider the Taylor expansion of the denominators in Eqs.~\eqref{eqn:coefficientSource}-\eqref{eqn:coefficientDrain}
\[
\frac{1}{\frac{E}{2J} - \cos\left(\frac{\pi k}{M+1}\right)} = -\sum_{p=0}^\infty \left(\frac{E}{2 J}\right)^p \sec^{p+1}\left(\frac{\pi k}{M+1}\right)
\]
obtaining
\begin{align}
    E \alpha_S &= -\frac{J'^2}{J(M+1)} \sum_{p=0}^\infty \left(\frac{E}{2 J}\right)^p \left(\beta_p^+ \alpha_S - \beta_p^-\alpha_D\right),\label{eqn:EalphaSource}\\
    E \alpha_D &= \frac{J'^2}{J(M+1)} \sum_{p=0}^\infty \left(\frac{E}{2 J}\right)^p \left(\beta_p^- \alpha_S -\beta_p^+\alpha_D\right),
    \label{eqn:EalphaDrain}
\end{align}
with
\begin{align}
    \beta_p^+ &= \sum_{k=1}^M \sin^2 \left(\frac{\pi k}{M+1}\right)
\sec^{p+1}\left(\frac{\pi k}{M+1}\right),
\label{eqn:summationSource}
\\
    \beta_p^- &= \sum_{k=1}^M (-1)^k \sin^2 \left(\frac{\pi k}{M+1}\right)\sec^{p+1}\left(\frac{\pi k}{M+1}\right)   \label{eqn:summationDrain}.
\end{align}
Note that for even $p$, $\beta_p^+ = 0$ and $\beta_p^- \ne 0$, while $\beta_p^- = 0$ and $\beta_p^+ \ne 0$ for odd $p$. To determine $\w_-$, we keep the first two terms in the summation over $p$, which can be evaluated analytically. Indeed, for even values of $M$,
\begin{align}
\beta_0^-&=\sum_{k=1}^M (-1)^k{\rm sec}\left(\frac{\pi k}{M+1}\right)-\sum_{k=1}^M (-1)^k\cos\left(\frac{\pi k}{M+1}\right)\nonumber\\
&=\sum_{k=1}^M (-1)^k{\rm sec}\left(\frac{\pi k}{M+1}\right)+1\nonumber\\
&=\sum_{k=0}^M (-1)^k{\rm sec}\left(\frac{\pi k}{M+1}\right)\nonumber\\
&=(-1)^{M/2} (M+1).
\label{eqn:summationBeta0Minus}
\end{align}
Above, we used the fact that the alternating finite cosine sum is $-1$ between the first and the second line; in the last line, we exploited a specific case of the alternating secant sum with odd power discussed in Ref.~\cite{WANG20071020}. Similarly, 
\begin{align}
\beta_1^+&=\sum_{k=1}^M (-1)^k{\rm sec}^2\left(\frac{\pi k}{M+1}\right)-\sum_{k=1}^M 1\nonumber\\
&=\sum_{k=0}^M (-1)^k{\rm sec}\left(\frac{\pi k}{M+1}\right)-(M+1)\nonumber\\
&=(M+1)^2-(M+1)=M (M+1),
\label{eqn:summationBeta1Plus}
\end{align}
where we used an additional result presented in Ref.~\cite{WANG20071020}. Keeping terms up to $p=1$ in Eqs.~\eqref{eqn:summationSource}-\eqref{eqn:summationDrain}, yields a homogeneous linear system in the two variables $\alpha_S,\alpha_D$. The slow-frequency $E=\w_-$ is obtained by requiring the discriminant of the linear system to be zero, yielding [upon insertion of \cref{eqn:summationBeta0Minus,eqn:summationBeta1Plus} into \cref{eqn:EalphaSource,eqn:EalphaDrain}]
\begin{equation}
    \w_- = \frac{J'^2}{J[1+M J'^2/(2J^2)]} \, .
\label{eqn:density-U0-even-correction-E}
\end{equation}
This more accurate analytical approximation for $\w_-$ is used to calculate the orange crosses in Figs.~\ref{fig5:std-transport-weak}(d)-(f).

\subsection{Chain length dependence}

\begin{figure}
    \centering
    \includegraphics[width=\linewidth]{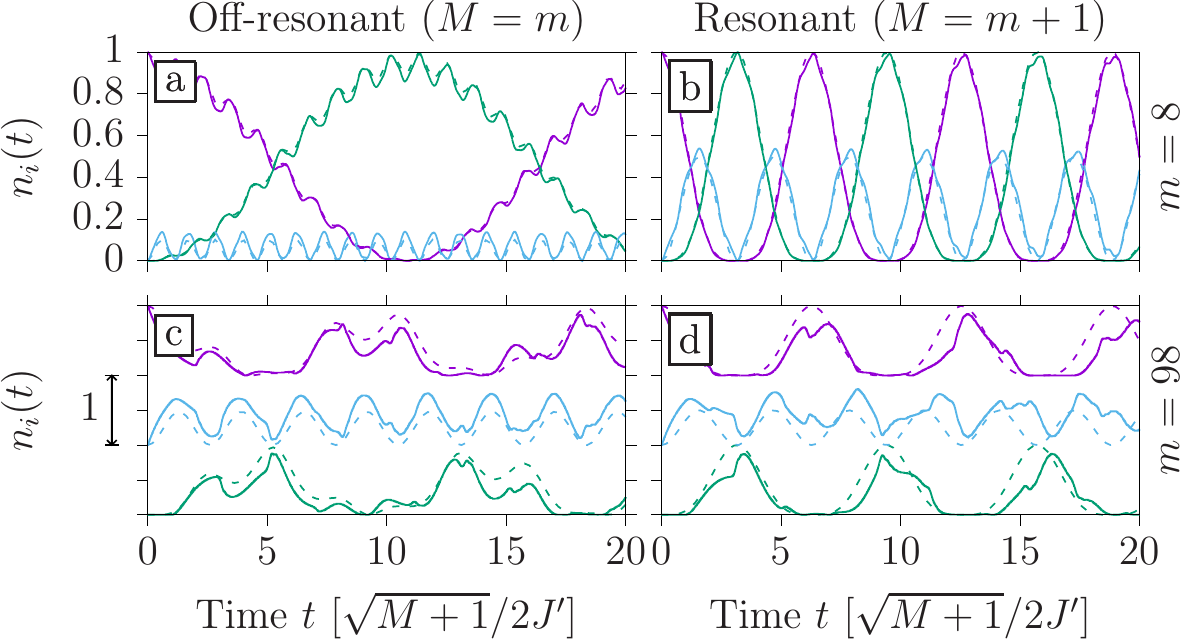}
    \caption{The time evolution of the source (purple), chain (blue), and drain (green) densities for a single-particle (thus $U=0$), $J'/J=0.1$, and $M = $ \textbf{(a)} 8, \textbf{(b)} 9, \textbf{(c)} 98, and \textbf{(d)} 99 sites, where an even (odd) number of sites corresponds to the off-resonant (resonant) case (colour online). The source and the chain curves have been shifted vertically in panels \textbf{(c)} and \textbf{(d)} for visualization purposes. The dashed lines correspond to the same analytic equations used to obtain the orange crosses of Figs.~\ref{fig5:std-transport-weak}(d)-(f),(j)-(l) [see Eqs.~(\ref{eqn:density-source-U0})-(\ref{eqn:density-drain-U0-even}) and the paragraphs containing them]. For a low number of sites, a strong parity effect in $M$ is observed. For a large number of sites, it is no longer easy to determine which of \textbf{(c)} and \textbf{(d)} corresponds to the off-resonant and resonant case by solely inspecting the dynamics in the chain. Time has been rescaled by $2 J'/\sqrt{M+1}$.}
    \label{fig:stdlargeMsamples}
\end{figure}

Here, we investigate the fate of the even/odd effect observed in \cref{sec:sourcetodrain} for $U=0$ and over large values of chain length $M$. As the system size increases, the level spacing for the chain's single-particle eigenvalues near zero energy vanishes [see Eq.~\eqref{eq:HTBDiag}], eventually eliminating the difference between even and odd parities. To estimate the critical size of the chain for which the parity difference becomes negligible, we consider the off-resonant case of \cref{app:even-chain}; within the approximation where only two states of the chain are taken into account, the frequencies characterizing the dynamics are given in \cref{eqn:even-chain-w-pm}. In the limit $M\to\infty$, from \cref{eq:betadef} we find $\beta \simeq 2\sqrt{2(M+1)}/\pi(J'/J) \gg 1$; it follows that $\w_+ \simeq \w_- \approx \frac{2 J'}{\sqrt{M+1}}$, the same value as  the splitting obtained in the resonant case at zero detuning, cf. \cref{eqn:resonantDensities} and the text that follows it. Moreover, as $\w_+$ approaches $\w_-$, the dimensionless parameter $\alpha=(\w_+ -\w_-)/(\w_+ +\w_-)$ decreases, thus increasing the occupation in the chain $N\propto 1-\alpha^2$. These results suggest that the difference between even and odd cases may become less marked as $M$ increases, with the dimensionless parameter controlling the crossover being $\tilde{\beta}=\sqrt{M+1}(J'/J)$; we expect that for $\tilde{\beta} \ll 1$, even and odd chains will display the markedly different behaviors described in Sec.~\ref{sec:sourcetodrain}. [Note that the smallness of $\tilde{\beta}$ ensures the validity of the more accurate estimate for $\omega_-$ in \cref{eqn:density-U0-even-correction-E}.]

We evaluate numerically the time evolution of the densities for different values of $\tilde\beta$. For $\tilde\beta\simeq0.3$
[see \cref{fig:stdlargeMsamples}(a) (even chain) and (b) (odd chain)],
the parity effect is well-marked. Indeed, the source/drain densities evolve slowly (with characteristic frequency $\w_-\ll \w_+$) in the off-resonant (even length) case as compared to the resonant (odd) case; accordingly, the density in the chain is small for even chains. As a result, the densities are well approximated by the analytical formulas of this Appendix. For longer chains with $\tilde\beta\simeq 1$, \cref{fig:stdlargeMsamples}(c) and (d), there is no qualitative difference in the chain occupation between even and odd chains. Interestingly, the analytical formulas still capture the main contributions to the dynamics, despite the increased contributions by other states further in energy from the band center. Finally, we observe that the dimensionless parameter determining the parity effect $\tilde{\beta}=\sqrt{M+1}(J'/J)$ is also associated with the perfect state transfer discussed in Ref.~\cite{wojcik2005unmodulated}, where the infidelity of the transfer scales as $\tilde{\beta}^2$ for $\tilde{\beta}\ll 1$. 


\begin{thebibliography}{69}
\bibitem [{foo()}]{footnote1}%
  \BibitemOpen
  \href@noop {} {}\bibinfo {note} {For a given number of particles, there is no
  dynamics when the distance between the boson stack or quantum soliton and the
  nearest boundary is smaller than $i_\text{pin}<\lceil N/2\rceil -1$, see
  Ref.~\cite{SilveriBeyondHardcore}.}\BibitemShut {Stop}%
\end{thebibliography}

%

\end{document}